\documentclass[12pt,bibliography]{article}
\usepackage{hyperref}
\usepackage{epsfig}
\usepackage{amsfonts} 
\usepackage{amsmath} 
\usepackage{color}

\def\a{\alpha}

\def\Or[#1]{{\text{O}}\left({#1}\right)}
\def\dotl[#1,#2]{\left\langle #1, #2 \right\rangle}
\def\dotlb[#1,#2]{[ #1, #2 ]}
\def\dotp[#1,#2]{(#1) \cdot (#2)}
\def\aff[#1,#2]{\hat{#1}(#2)}
\def\n4sym{{\cal N}=4 SYM}
\def\>{\rangle}
\def\<{\langle}
\def\weight[#1,#2,#3]{\{(#1),#2,#3\}}
\def\ads[#1]{$\text{AdS}_{#1}$}

\newcommand{\ba}{\begin{eqnarray}}
\newcommand{\ea}{\end{eqnarray}}

\newcommand{\be}{\begin{equation}}
\newcommand{\ee}{\end{equation}}  
\newcommand{\bi}{\begin{itemize}}
\newcommand{\ei}{\end{itemize}}

\newcommand{\aslash}[1]{\,\,{\raise.15ex\hbox{/}\mkern-12mu #1}}
\newcommand{\bslash}[1]{\,\,{\raise.15ex\hbox{/}\mkern-9mu #1}}

\renewcommand{\bar}{\overline}
\renewcommand{\tilde}{\widetilde}
\renewcommand{\hat}{\widehat}

\newcommand\lrpar{\raise .8ex\hbox{$^\leftrightarrow$} \hspace{-9pt}
\partial}
\newcommand\lpar{\raise .8ex\hbox{$^\leftarrow$} \hspace{-9pt}
\partial}
\newcommand\rpar{\raise .8ex\hbox{$^\rightarrow$} \hspace{-9pt}
\partial}
\newcommand\lrd{\raise .8ex\hbox{$^\leftrightarrow$} \hspace{-9pt}
\nabla}

\newcommand{\gsim}{\lower.7ex\hbox{$\;\stackrel{\textstyle>}{\sim}\;$}}
\newcommand{\lsim}{\lower.7ex\hbox{$\;\stackrel{\textstyle<}{\sim}\;$}}

\let\a=\alpha  \let\g=\gamma \let\d=\delta 
    \let\k=\kappa
\let\l=\lambda  \let\n=\nu  
     
  \let\D=\Delta  
    
    \let\G=\Gamma

\renewcommand{\ba}{\begin{eqnarray}}
\renewcommand{\ea}{\end{eqnarray}}
\newcommand{\bea}{\begin{eqnarray}}
\newcommand{\eea}{\end{eqnarray}}

\begin{document}

\begin{titlepage}

\begin{center}
\vspace{1cm}

{\Large \bf  Spinning AdS Propagators}

\vspace{0.8cm}

{\bf Miguel S. Costa, Vasco Gon\c calves, Jo\~ao Penedones}

\vspace{1cm}

{\it  Centro de F\'\i sica do Porto \\
Departamento de F\'\i sica e Astronomia\\
Faculdade de Ci\^encias da Universidade do Porto\\
Rua do Campo Alegre 687,
4169--007 Porto, Portugal}

\end{center}
\vspace{1cm}

\begin{abstract}
We develop the embedding formalism to describe symmetric traceless tensors in Anti-de Sitter space. We use this formalism to construct the bulk-to-bulk  propagator of massive spin $J$ fields and  check that it has the expected short distance and massless limits.
We also find a split representation for the bulk-to-bulk propagator, by writing it as an integral over the boundary of the product of two bulk-to-boundary propagators.
We exemplify the use of this representation with the computation of the conformal partial wave decomposition of Witten diagrams. In particular, we determine the Mellin amplitude associated to AdS graviton exchange between minimally coupled scalars of general dimension, including the regular part of the amplitude.
\end{abstract}

\bigskip
\bigskip

\end{titlepage}

\tableofcontents

\newpage

\section{Introduction}

Higher spin fields play an important role in all known examples of the AdS/CFT duality \cite{Maldacena}. In the case of the ${\cal N}=4$ SYM there are massive string states in AdS, which must be taken
into account at finite 't Hooft coupling $\lambda$, and whose effect appears at infinite coupling in the form of $1/\sqrt{\lambda}$ corrections. More recently there has been a resurgence 
of  higher spin  gauge theories in $AdS_4$ \cite{Vasiliev:1990en,Vasiliev:1992av,Vasiliev:1995dn,Vasiliev:1999ba}, 
conjectured to be dual to three-dimensional CFTs like the $O(N)$ vector model \cite{KlebanovPolyakov}, the Gross-Neveu model \cite{Sezgin:2003pt} or certain 
large $N$ Chern-Simons theories \cite{Leigh:2003gk}.  
In all these recent cases, computations involving  AdS higher spin fields pose additional technical challenges. The goal of this paper is to develop a formalism
to deal with tensor fields in AdS, that makes computations almost as simple as those with scalar fields.

More specifically, we shall develop the embedding formalism for treating massive  symmetric traceless AdS tensor fields with $J$ indices (or spin $J$ fields, for short), but some of the methods here developed should  
be extendable to antisymmetric tensors or mixed symmetry tensors. The basic idea of the embedding formalism is that fields in Euclidean $AdS_{d+1}$ space, or their $CFT_d$ dual operators, 
can be expressed in terms of fields in an embedding Minkowski  space $\mathbb{M}^{d+2}$. The action of the $AdS_{d+1}$ isometry group, or of the conformal group $SO(d+1,1)$,  can then
be realised as the group of linear Lorentz transformations. This fact has been explored in many places in the literature to simplify computations, including
computations of correlation functions of higher spin fields \cite{ourDIS,Weinberg:2010fx,SpinningCC}, of conformal blocks for external operators with spin \cite{Costa:2011dw,SimmonsDuffin:2012uy}, and of Witten diagrams to derive Feynman rules in Mellin space \cite{JPMellin,PaulosMellin,NaturalMellin}, to name a few. 

We shall start, in Section \ref{EmbeddingFormalism}, by introducing the 
basic definitions  that allow us to describe AdS fields in the embedding formalism, including 
the representation of differential operators such as the Laplacian. As a first application of the formalism, we compute in Section  \ref{SpinJPropagators}
the bulk-to-bulk propagator of a massive spin $J$ field in  AdS. 
Explicit expressions for the scalar and spin 1 cases are known for a long time, while the expressions for the massive spin 2 and p-form cases are known only more recently \cite{Naqvi:1999va}.
The new propagator has the required short distance behaviour derived in \cite{Giombi:2013yva} using zeta function regularisation. It also reproduces the known form of the
vector propagator, as well as the traceless part of the graviton propagator in the massless limit, as given in \cite{D'Hoker:1999jc}.  
In general it is known that the bulk-to-bulk propagator is closely related to the product of two
bulk-to-boundary propagators integrated over a common boundary point \cite{Leonhardt:2003qu,Leonhardt:2003sn}.
In Section \ref{SplitRep} we make this relation precise by deriving a split representation for spin $J$ propagators in AdS.
We also consider the case of the graviton propagator whose split representation has generated some discussion in the literature \cite{Balitsky:2011tw,Giecold:2012qi}.  
Taking one of the bulk points to the boundary, the spin $J$  propagator also defines the bulk-to-boundary propagator for this field. 
As an application of this result we derive in Section  \ref{OPECubicAdScoupling}  the 
relation between the AdS local coupling of two scalar fields and one spin $J$ field, and the 
OPE coefficient of the  dual CFT operators. 
Finally, in Section \ref{CPWdecomposition} we make use of this split representation  to derive the conformal block
expansion  of  four-point correlation functions computed via  Witten diagrams.
In particular, we determine the Mellin amplitude associated to AdS graviton exchange between minimally coupled massive scalars in general spacetime dimension.
A number of  technical computations are left to   appendices.

\section{Embedding formalism for AdS}
\label{EmbeddingFormalism}
In this paper we consider tensor fields in  Euclidean $(d+1)$-dimensional Anti de Sitter space AdS$_{d+1}$.
Obviously this is just the  $(d+1)$-dimensional hyperbolic space.
Our expressions can be Wick-rotated to Minkowski signature, provided one is careful with the $i\epsilon$
prescription (see \cite{Mythesis,ourSW} for some details).
In this section we introduce notation and develop the 
embedding formalism to treat tensor fields in AdS$_{d+1}$. We shall see how the use
of this formalism simplifies computations considerably, making conformal invariance
manifest at all time, just like for 
tensor fields in $d$-dimensional CFTs.

\begin{figure}[t!]
\begin{centering}
\includegraphics[scale=0.4]{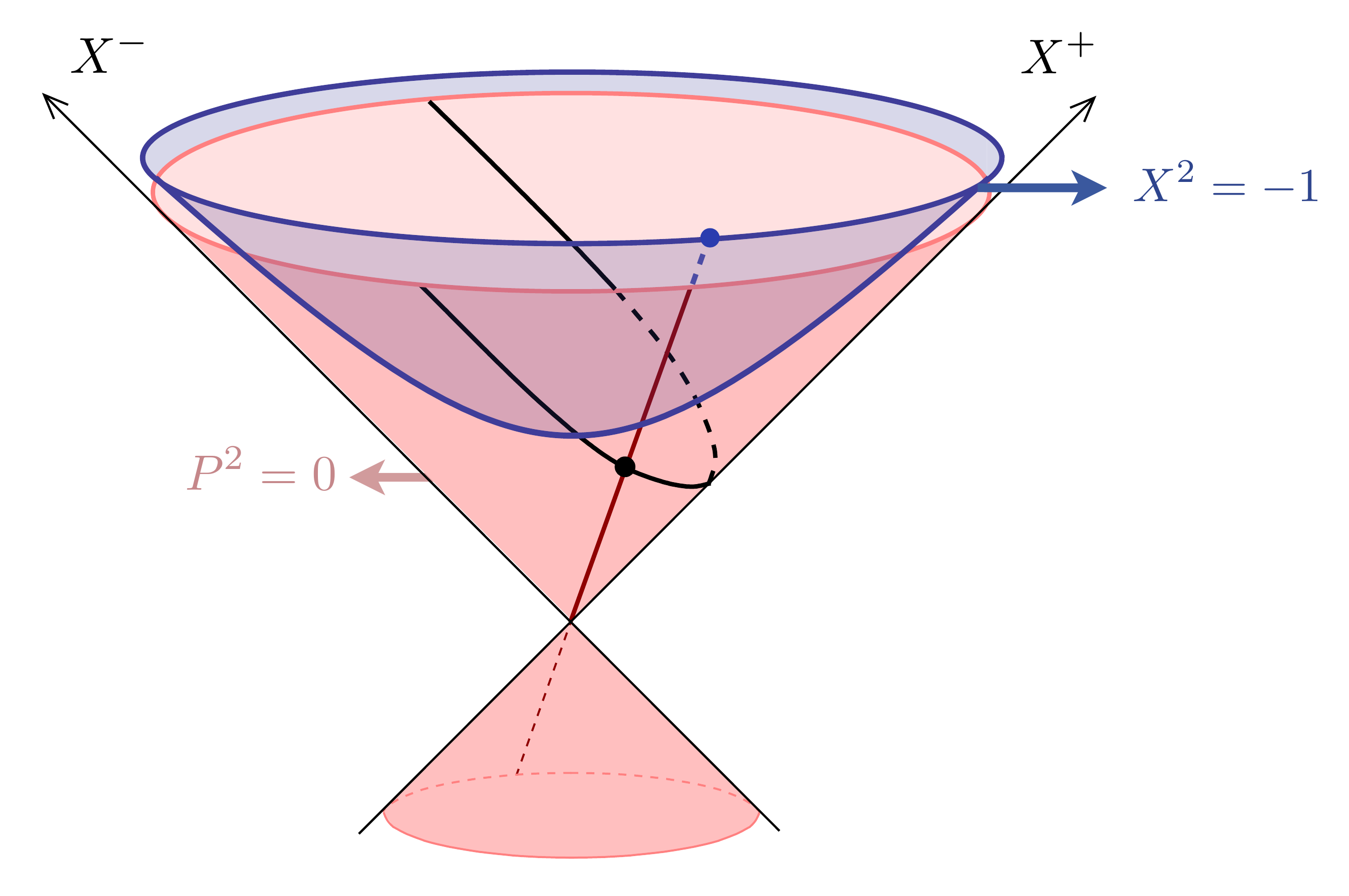}
\par\end{centering}
\caption{\label{fig:AdS}
Euclidean AdS and its boundary in the embedding space. This picture shows the $AdS_2$ surface $X^2=-1$ and the identification of a boundary point (in blue) with 
a light ray (in red) of the light cone $P^2=0$, which intersects  the Poincar\'e section on a (black) point.}
\end{figure}

Euclidean AdS$_{d+1}$ space can be defined by the set of future
directed unit vectors,
\be
X^{2}=-1\,,\ \ \ \ \ \ \ \ \ \ \ \ X^0>0\,,
\ee
in $(d+2)-$dimensional Minkowski space $\mathbb{M}^{d+2}$. As it is well known, 
the isometry group of AdS$_{d+1}$ is the $d$-dimensional conformal 
group $SO(d+1,1)$. This group acts linearly on the embedding space $\mathbb{M}^{d+2}$,
and its action is interior to points on the hyperboloid $X^{2}=-1$.
A simple example is that of  AdS$_{d+1}$ written in Poincar\'e coordinates $x^\mu= (z, y^a)$, with $y$ a $d$-dimensional vector. 
In this case AdS points are parameterized as
\be
X=\frac{1}{z}\left(1, z^2 + y^2, y^a\right)\label{eq:PoincareCoordinates},
\ee
where we used light cone coordinates 
\be
X^A=\left(X^+,X^-,X^a\right),
\ee
with metric 
\be
X\cdot X = \eta_{AB} X^A X^B = -X^+ X^- + \delta_{ab}X^a X^b\,.
\ee
Here and below, we use capital letters to denote embedding space indices in $\mathbb{M}^{d+2}$, lower case 
letters to denote indices in $\mathbb{R}^{d}$, and greek letters to denote AdS$_{d+1}$ indices.

AdS boundary points can be obtained by sending some of the $X$ coordinates to infinity. In this limit the hyperboloid approaches 
the light cone, so that  a given specific point at infinity in the hyperboloid approaches one light ray. This allows for the identification of the AdS
boundary with light rays, according to
\be
P^2=0\,,\ \ \ \ \ 
P\equiv \lambda P\,,\ \ \ \ \ \lambda \in \mathbb{R}\,.
\ee
For example, for the Poincar\'e patch considered above, boundary points are parameterised by
\be
P=\left(1, y^2, y^a\right).
\ee
Figure \ref{fig:AdS} summarizes  this geometric picture.

We wish to establish the relation between fields in AdS$_{d+1}$ and $\mathbb{M}^{d+2}$.
In particular, here we will consider  traceless symmetric tensors. Let us then consider a traceless symmetric tensor
of $\mathbb{M}^{d+2}$ with components $H_{A_{1}\dots A_{J}}(X)$, defined on the surface $X^{2}=-1$ and  transverse to this surface,
\begin{equation}
X^{A_{1}}H_{A_{1}\dots A_{J}}(X)=0\,.
\label{TransverseCondition}
\end{equation}
This defines a tensor in AdS$_{d+1}$, whose components are simply obtained by the projection
\be
h_{\mu_1\dots \mu_J} = \frac{\partial X^{A_1}}{\partial x^{\mu_1}}\cdots  \frac{\partial X^{A_J}}{\partial x^{\mu_J}}\,H_{A_{1}\dots A_{J}}(X)\,.
\label{projection}
\ee
The extension of the embedding tensor $H(X)$ away from the AdS$_{d+1}$ submanifold $X^{2}=-1$ is not physical. On one hand, this means that
components of the tensor that are transverse to the hyperboloid, i.e. of the type
\be
H_{A_{1}\dots A_{J}}(X)= X_{\left(A_1\right.} \Psi_{\left. A_{2}\dots A_{J}\right)}(X)
\,.
\ee
are unphysical. Indeed, these components, which do not satisfy the transverse condition (\ref{TransverseCondition}), have vanishing projection to AdS$_{d+1}$.
On the other hand, it also means that whenever we take a derivative in the embedding space, that derivative can only be tangent to the AdS$_{d+1}$
submanifold.

We wish to have a more economical
way of encoding AdS$_{d+1}$ tensors, without having to deal with all the indices and constraints arising  from the  linear realization of the
$SO(d+1,1)$ symmetry.
Let us first recall how this can be achieved in the case of $\mathbb{R}^{d}$ tensors, extensively discussed in \cite{SpinningCC}.
In this case a symmetric traceless tensor with components $F_{A_1\cdots A_J}(P)$ is defined on the light-cone $P^2=0$ of the 
embedding space with the requirement that $F(\lambda P) = \lambda^{-\Delta} F(P)$, for $\lambda>0$, where $\Delta$ is the conformal dimension. This 
tensor can be encoded in the polynomial 
\begin{equation}
F(P,Z ) = Z^{A_1}\dots Z^{A_J} F_{A_1\cdots A_J}(P)\,,
\end{equation}
where $Z^2=0$ encodes the traceless condition. To be tangent to the light-cone $P^2=0$ the embedding tensor must satisfy
 $P^{A_1}F_{A_1\cdots A_J}=0$, which can be implemented  by requiring 
 $F(P,Z +\a P)=F(P,Z )$ for any  $\a$.
In addition, we can impose  
 the orthogonality condition $P\cdot Z=0$ because $F_{A_1\cdots A_J} = P_{(A_1}\Psi_{A_2\cdots A_J)}$ has vanishing projection into physical $\mathbb{R}^d$ tensors. 
 Moving to the case of AdS symmetric traceless  tensors $H_{A_{1}\dots A_{J}}(X)$, defined on the submanifold $X^2=-1$, they can be encoded by  
 $(d+1)$-dimensional polynomials as
\begin{equation}
H(X,W)=W^{A_{1}}\dots W^{A_{J}}H_{A_{1}...A_{J}}(X)\,,
\label{polynomial}
\end{equation}
where $W^{2}=0=X\cdot W$. The traceless condition allows one to restrict the polynomial to the submanifold $W^2=0$, and the transverse
condition allows for the further restriction $X\cdot W=0$. In sum, a 
symmetric traceless tensor can be fully encoded by a polynomial $H(X,W)$
defined on the submanifold $X^{2}+1=W^{2}=X\cdot W=0$.
To recover the AdS tensor from a given polynomial we define the operator
\begin{align}
K_{A}= &\ \frac{d-1}{2} \left(\frac{\partial}{\partial W^{A}}+X_{A} \left( X\cdot\frac{\partial}{\partial W} \right)\right)+
\left(W\cdot\frac{\partial}{\partial W}\right)\frac{\partial}{\partial W^{A}}+
 \label{Projector}
\\
 & +X_{A}\left( W\cdot\frac{\partial}{\partial W}\right) \left(X\cdot\frac{\partial}{\partial W} \right)
 -\frac{1}{2}\,W_{A}\left(\frac{\partial^{2}}{\partial W\cdot\partial W}+ \left(X\cdot\frac{\partial}{\partial W} \right)\left(X\cdot\frac{\partial}{\partial W}\right)\right)\,.
\nonumber
\end{align}
We constructed this second order differential operator such that
it is interior with respect to the submanifold $X^{2}+1=W^{2}=X\cdot W=0$ (i.e. its action on a function only depends on the value of the function on this submanifold).
Moreover,  it is transverse ($X^{A}K_{A}=0$), symmetric ($K_AK_B=K_BK_A$) and traceless ($K_{A}K^{A}=0$), so that its action on any polynomial of $W$ will define a transverse
symmetric traceless AdS tensor. 
To be precise, it acts as a projector since
\be
\frac{1}{J!\left(\frac{d-1}{2}\right)_J}\,K_{A_1}\dots K_{A_J} W^{B_1}\dots W^{B_J} 
=G_{\left\{A_1\right.}^{\ \ \ B_1}\dots G_{\left.A_J\right\}}^{\ \ \ \,B_J}
\,,
\ee
where 
\be
G_{AB}=\eta_{AB}+X_AX_B\,,
\label{AdS_metric}
\ee is the induced AdS metric (and therefore a projector).
Our convention for the index symmetrization  is normalized according to
\be
G_{\left\{A_1\right.}^{\ \ \ B_1}\dots G_{\left.A_J\right\}}^{\ \ \ \,B_J}
 =  \frac{1}{J!}\sum_\pi G_{A_{\pi_1}}^{\ \ \ B_1}\dots G_{A_{\pi_J}}^{\ \ \ \,B_J} - {\rm traces}\,,
\label{sim_convention}
\ee
where the sum is over all permutations of the $A$ indices and we subtract the traces using the AdS metric $G_{AB}$.
In particular, notice that acting on the polynomial 
 (\ref{polynomial}), the projector $K_A$ simplifies to
\begin{align}
K_{A}= &  \left(
\frac{d-1}{2}
+W\cdot\frac{\partial}{\partial W}\right)\frac{\partial}{\partial W^{A}}
\end{align}
because the tensor $H_{A_{1}\dots A_{J}}$ is already traceless and transverse.
It is then straightforward to show that the components of the symmetric and traceless  AdS tensor in (\ref{polynomial}) can be 
recovered from its polynomial via
\begin{equation}
H_{A_{1}\dots A_{J}}(X)=\frac{1}{J!\left(\frac{d-1}{2}\right)_J}\,K_{A_{1}}\dots K_{A_{J}}H(X,W)\,,
\label{eq:PolProj}
\end{equation}
where $(a)_J=\Gamma(a+J)/\Gamma(a)$ is the Pochhammer symbol. 
Thus from now on we will work with polynomials that uniquely determine AdS symmetric traceless tensors.

Our main goal in this paper is to construct in a systematic way and in full generality the AdS propagator for a massive spin $J$   field. 
This means that we need to define an embedding differential operator that computes the AdS covariant derivative.
Acting on symmetric traceless tensors encoded in polynomials of $W$, as in (\ref{polynomial}), the embedding differential operator that does the job is 
\begin{align}
\nabla_{A}=\frac{\partial}{\partial X^{A}}+X_{A}\left( X\cdot\frac{\partial}{\partial X} \right)+W_{A}\left( X\cdot\frac{\partial}{\partial W}\right).
\label{Derivative}
\end{align}
As necessary, this operator is interior to the submanifold $X^{2}+1=W^{2}=X\cdot W=0$
and transverse  ($X^{A}\nabla_{A}=0$). With the help of this differential operator and
of the projector (\ref{Projector}), we can compute the divergence of a tensor by 
\begin{equation}
(\nabla\cdot H)(X,W)=
\frac{1}{J\left(\frac{d-3}{2} +J\right)}\,\nabla\cdot K\,H(X,W)\,.
\end{equation} 
The left hand side of this equation is 
the polynomial  whose projection to AdS gives the 
divergence $D_{\mu_1} h^{\mu_{1}}_{\ \ \mu_2...\mu_{J}}$.  Instead, 
in the right hand side we freed first one embedding index acting with $K_A$, and then contracted it with the embedding differential operator $\nabla_{A}$.
\footnote{On the submanifold $X^{2}+1=W^{2}=X\cdot W=0$, the order of the operators $K_A$ and $\nabla^A$ is not important because  $\nabla\cdot K=K\cdot \nabla$.}

The Laplacian
of a tensor field in AdS, $\nabla^2H$, can be simply recovered from the  polynomial
\begin{equation}
\left(\nabla^2H\right) (X,W) = \nabla\cdot\nabla\,H(X,W)\,,
\end{equation}
which, after projection to AdS,  computes $D^{\nu}D_{\nu}h_{\mu_1...\mu_{J}}$.

The embedding space can also be used to compute covariant derivatives of more general tensors (with open indices). Given an embedding tensor $T_{A_1\dots A_n}$ obeying the transversality condition (\ref{TransverseCondition}), its covariant derivative is simply given by 
\be
\nabla_B T_{A_1\dots A_n}(X)= G_B^{\ \, C} G_{A_1}^{\ \ C_1} \dots G_{A_n}^{\ \ C_n}\, \frac{\partial}{\partial X^C} \,T_{C_1\dots C_n}(X)\,,
\label{generalnabla}
\ee
where the projector $G_{B}^{\ \,C}$ is the AdS metric given in (\ref{AdS_metric}).

%

\section{AdS propagators of spinning particles}
\label{SpinJPropagators}

Let us first recall some basic results on particles with spin 1 and 2.
 A massive spin 1 particle is described by the Euclidean   action
\be
\int_{AdS} d^{d+1}x\sqrt{g}\left[ \frac{1}{2}(D_\mu A_\nu)^2 -
\frac{1}{2}(D^\mu A_\mu)^2 +  
\frac{1}{2}M^2 A^\mu A_\mu  -A_\mu j^\mu \right],
\ee
where $D_\mu$ is the AdS covariant derivative and $j^\mu$ is a classical source.
This action gives rise to the Proca equation
\be
D^2 A_\mu -
D_\mu(D^\nu A_\nu) -M^2 A_\mu =-j_\mu\,.
\ee
Taking the divergence of this equation in the absence of source, we conclude that $D^\nu A_\nu=0$. Thus,  in the absence of source, the Proca equation is equivalent to 
$D^2 A_\mu=M^2 A_\mu$ and $D^\nu A_\nu=0$.

 A massive spin 2 particle is described by the Euclidean  action \cite{Fierz:1939ix, 
 Higuchi:1986py,Naqvi:1999va}
 \begin{align}
\int_{AdS} d^{d+1}x\sqrt{g}\, 
&\left[ 
\frac{1}{2}(D_\mu h_{\nu\a})^2
-\frac{1}{2}(D_\mu h)^2
+ D^\mu h_{\mu\nu} D^\nu h
-D_\mu h_{\nu\a} D^\a h^{\nu\mu}
\right.
\label{spin2action}\\&
\ \ \left.+d(h_{\mu\nu})^2
+\frac{d}{2}h^2+
\frac{1}{2}(M^2+2) (h_{\mu\nu}^2- h^2)  
-T^{\mu\nu}h_{\mu\nu} \right],
\nonumber
\end{align}
where  $T^{\mu\nu}$ is a classical source and $h=g^{\mu\nu}h_{\mu\nu}$ is the trace of the field $h_{\mu\nu}$.
This action can be obtained from the Einstein-Hilbert action for the metric $g_{\mu\nu}+h_{\mu\nu}$ in the presence of a negative cosmological constant equal to $-d(d-1)/2$, by expanding to quadratic order in the metric fluctuation $h_{\mu\nu}$, and adding the Fierz-Pauli mass term $\frac{1}{2}(M^2+2) (h_{\mu\nu}^2- h^2)$ to the lagrangian. 
The equation of motion can be written as \cite{Naqvi:1999va}
\begin{align}
 &(D^2   -M^2)h_{\mu \nu }
 -  D_{\mu } D^\sigma h_{ \sigma \nu  }-  D_{\nu} D^\sigma h_{\mu  \sigma  }-2g_{\mu  \nu} h   
 \label{eomspin2massive}\\
 &+\bigg(D_{\mu } D_{\nu}-\frac{M^2+2}{d-1} \,g_{\mu  \nu}\bigg)h   =
-T_{\mu\nu}+\frac{1}{d-1} \,g_{\mu  \nu} T_{\sigma}{}^\sigma{} \,,
\nonumber
\end{align}
where we recall that  the AdS radius is  1 in our units.
By taking the trace and the divergence of this equation in the absence of source, we can derive  
\be
D^2 h_{\mu\nu}=M^2 h_{\mu\nu}\,,\ \ \ \ \ \ \
D^\mu h_{\mu\nu}=0\,,\ \ \ \ \ \ \
h_{\mu}{}^ \mu{}  =0\,.
\ee

The generalization of these equations to higher spin is more  involved. Starting from spin 3 one must either introduce non-local terms or auxiliary fields \cite{Bouatta:2004kk, Francia:2007ee, Francia:2010ap}. However, on-shell, these equations always reduce to
\be
D^2 h_{\mu_1\dots \mu_J}=M^2 h_{\mu_1\dots \mu_J}\,,\ \ \ \ \ \ \
D^{\mu } h_{\mu \mu_2\dots \mu_J}=0\,,\ \ \ \ \ \ \
h^{\mu}_{\ \ \mu \mu_3\dots \mu_J}  =0\,.
\label{eq:FP}
\ee
This will be enough for our purposes, since it determines all poles associated with propagating degrees of freedom.

\subsection{Bulk-to-bulk propagator}
\label{Sec:PropagatorJ}

To construct the  bulk-to-bulk propagator of a spin $J$ field between points $X_1$ and $X_2$, respectively  with polarization 
vectors $W_1$ and $W_2$,  we need to consider polynomials of degree $J$ in both $W_1$ and $W_2$ that can be constructed from the 
three possible scalar products $W_{1}\cdot W_{2}$, $X_{1}\cdot W_{2}$ and $X_{2}\cdot W_{1}$. The coefficient of each term can
be a generic function of the chordal distance $u=-1-X_{1}\!\cdot\! X_{2}$. Thus we write with full generality
\begin{equation}
\Pi_{\Delta,J}(X_{1},X_{2},W_{1},W_{2})=\sum_{k=0}^{J}(W_{12})^{J-k}\big((W_{1}\cdot X_{2})(W_{2}\cdot X_{1})\big)^{k}g_{k}(u)\,,
\label{eq:propagatorJgk}
\end{equation}
where we introduced the notation $W_{12}= W_{1}\cdot W_{2}$.

To see how this formalism relates to the more conventional treatment, let us consider the simple case of $J=1$ and arbitrary dimension $\Delta$. In this case we have
\begin{equation}
\Pi(X_{1},X_{2},W_{1},W_{2})= W_{12} \,g_{0}(u)
+(W_{1}\cdot X_{2})(W_{2}\cdot X_{1})\,g_{1}(u)\,.
\label{eq:propagatorJ=1}
\end{equation}
Next we should act with the projector operator (\ref{Projector}) to recover the components of the propagator as an embedding tensor
\begin{align}
\Pi_{A,B}(X,Y)= &\,\big(  \eta_{AB} +X_{A}X_{B} + Y_{A}Y_{B}  - (1+u ) X_{A}Y_{B} \big)
 \,g_{0}(u) 
 \nonumber\\
&+ 
\big(X_B  -(1+u)  Y_B \big)
\big(Y_A  -(1+u)  X_A \big)
\,g_{1}(u)\,.
\end{align}
Finally we can project to some AdS coordinate system using (\ref{projection}). The terms proportional to $X_A$ or $Y_B$ are then seen to have a vanishing  projection.
The result can be  expressed, in terms of the usual tensor structures constructed
from derivatives of the chordal distance between both points, as
\begin{equation}
\Pi_{\mu,\nu}(x,y)=  - \frac{\partial^2 u}{\partial x^\mu \partial y^\nu} \,g_{0}(u)  +  \frac{\partial u}{\partial x^\mu } \frac{\partial u}{\partial y^\nu}
\,g_{1}(u)\,.
\label{eq:AdSpropagatorJ=1}
\end{equation}

Let us return to the problem of finding the general form of the functions $g_{k}(u)$ in (\ref{eq:propagatorJ}). 
Based on a similar  analysis in flat space that we include in appendix \ref{Ap:flatspaceOmega}, there is an alternative way other than (\ref{eq:propagatorJgk})
of writing the propagator that turns out to simplify the computation,
\begin{equation}
\Pi_{\Delta,J}(X_{1},X_{2},W_{1},W_{2})=\sum_{k=0}^{J}(W_{12})^{J-k}\big((W_{1}\cdot\nabla_{1})(W_{2}\cdot\nabla_{2})\big)^{k}f_{k}(u)\,.
\label{eq:propagador AdS}
\end{equation}
The equivalence of expressions (\ref{eq:propagatorJgk}) and (\ref{eq:propagador AdS}) relates the functions $g_{k}(u)$ and $f_{k}(u)$ through
\begin{equation}
g_{k}(u)=\sum_{i=k}^{J}\left(-1\right)^{i+k}\left(\frac{i!}{k!}\right)^{2}\frac{1}{\left(i-k\right)!} f_{i}^{\left(i+k\right)}(u)\,,
\label{eq:equacao entre g e f}
\end{equation}
where $f_i^{(k)}(u)=\partial_u^k f_i (u) $ denotes
the $k$-th derivative of $ f_i (u)$.

The equations  for the bulk-to-bulk propagator of a massive spin $J$ field are given by
\begin{align}
\left(\nabla_{1}^{2}-\Delta(\Delta-d)+J\right)\Pi_{\Delta,J}(X_{1},X_{2},W_{1},W_{2}) & =-\delta(X_{1},X_{2})\, (W_{12})^{J}+\dots\,,
\label{eq:Laplacian}\\
\nabla_{1}\cdot K_{1}\,\Pi_{\Delta,J}(X_{1},X_{2},W_{1},W_{2}) & =\dots	\,,
\label{eq:divergence}
\end{align} 
where we wrote the mass squared  in (\ref{eq:FP}) in AdS units as $M^2=\Delta(\Delta-d)-J$, such that $\D$ is the dimension of the dual operator.
In these equations, the dots represent   local source terms that are not important for the propagating degrees of freedom. As we shall see, they only change the propagator by contact terms. 
Since we will reproduce known formulae for lower spin fields, and also to make explicit our normalisation of the delta
function singularity in the propagator equation, it is helpful to write these two equations
in terms of  components of the physical tensors.
A mechanical computation shows that acting with the projector (\ref{Projector}) we obtain the familiar equations
\begin{align}
\left(D_1^2-\Delta(\Delta-d)+J\right)\Pi_{\mu_1\dots\mu_J,\nu_1\dots\nu_J}(x_1,x_2) & =
 - g_{\mu_1\left\{\nu_1\right.}\cdots g_{|\mu_J|  \left. \nu_J\right\}}\,\delta(x_{1},x_{2})+\dots\,,
\label{eq:Laplacian1}\\
D_1^{\mu_1} \Pi_{\mu_1\dots\mu_J,\nu_1\dots\nu_J}(x_1,x_2) &=\dots\,,
\label{eq:divergence1}
\end{align} 
where $D_1$ is the covariant derivative acting on functions of  $x_1$ and we use the same convention for
index symmetrization as given in (\ref{sim_convention})

The simplicity brought by the formalism can now be  appreciated by the action of the Laplacian on our ansatz 
(\ref{eq:propagador AdS}). One obtains for the   propagator equation (\ref{eq:Laplacian}) the expression
 \begin{align}
&\!\! \! (W_{12})^{J}\delta \!\left(X_1,X_2\right)+\dots=
\sum_{k=0}^{J}\left(W_{12}\right)^{J-k-1} 
\bigg[
2(J-k)\left(W_{1}\cdot\nabla_{1}\right)^{k+1}\left(W_{2} \cdot \nabla_{2}\right)^{k} \left(X_{1} \cdot  W_{2}\right) \Big.+
\label{eq:lapla}\\
&\!\! \! 
\Big.
 W_{12}\big((W_{1} \cdot \!\nabla_{1})(W_{2} \cdot \!\nabla_{2})\big)^{k}
\Big( \!u \left(2+u\right)\!\partial_u^2+ (d+1) \left(1+u\right)\!\partial_u+k\left(2+k-2J\right)-\D(\D-d)\!\Big)\!\bigg]f_{k}\,.
\nonumber
\end{align}
Although it is not explicit, this equation is actually symmetric under exchange of points 1 and 2. In fact, 
the  term $(X_{1}\cdot W_{2})f_{k}\left(u\right)$, arising from the first line,
can be written as
\be
-W_{2}\cdot\nabla_{2}\int^{u}du'f_{k}(u')\,,
\ee
so the tensor structure of this term can be obtain from that of the second line in (\ref{eq:lapla}) by setting $k\to k+ 1 $.
Further simplification is achieved by using instead the $k$-th derivative of $f_k$, since in (\ref{eq:equacao entre g e f}) 
there are always at least $k$ derivatives of $f_k$. Thus, using the shorthand notation $h_k=f_k^{\left(k\right)}$, (\ref{eq:lapla}) becomes 
\begin{align}
&\Big(u (2+u)\partial_u^2+(d+1 )(1+u)\partial_u
 -\Delta (\Delta-d)\Big)h_{0}=0 \,,
\label{eq:RecursionRel}\\
&\Big(u (2+u)\partial_u^2+(d+1+2k)(1+u)\partial_u+2k(k-J+1)-\Delta (\Delta-d)\Big)h_{k}=2(J+1-k)h_{k-1} \,,
\nonumber
\end{align}
 for $k=0$ and $k>0$, respectively.
The former is nothing more than the equation for the scalar propagator.  
Solving the latter for $J$ up to 7 we found a recurrence relation for $h_{k}$ in terms of $h_{k-1}$ and $h_{k-2}$.
Inspired by this result we conjecture that the general solution is defined recursively by
\begin{equation}
h_{k}=c_{k} \Big((d-2k+2J-1)\left(\left(d+J-2\right)h_{k-1}+\left(1+u\right)h'_{k-1}\right)+\left(2-k+J\right)h_{k-2}\Big)\,,
\label{eq:solution}
\end{equation}
where 
\begin{equation}
c_{k}=-\frac{\left(1+J-k\right)}{k\!\left(d+2J-k-2\right)\!\left(\Delta+J-k-1\right)\!\left(d-\Delta+J-k-1\right)} \,,
\label{eq:ak}
\end{equation}
and 
\begin{equation}
h_0\!\left(u\right)=
\frac{\Gamma\!\left(\Delta\right)}{2\pi^{\frac{d}{2}}  \Gamma\!\left(\Delta+1-\frac{d}{2}\right)}
\left( 2u \right)^{-\Delta} \!\!\ _2F_1\!\left(\Delta,\Delta+\frac{1-d}{2},2\Delta-d+1,-\frac{2}{u}\right).
\label{eq:h0} 
\end{equation}
The normalization of $h_0$ is fixed by the $\d$-function source in the propagator equation.
The equation for the divergence (\ref{eq:divergence}) was checked to hold for $J$ up to 6. Previous results for propagators in AdS were confirmed for $J=1$ and $J=2$, as we now discuss
\cite{Allen:1985wd,Naqvi:1999va,Leonhardt:2003sn,Leonhardt:2003qu}. 

\subsubsection{Spin 1}

In the case $J=1$ explicitly considered above, using
(\ref{eq:equacao entre g e f}) and $f_{i}^{\left(i+k\right)}=h_i^{\left(k\right)}$, it is simple to see that the functions of the chordal distance that multiply the different tensor structures, as 
described by (\ref{eq:AdSpropagatorJ=1}), are given by
\begin{align}
&g_0(u)=  (d-\Delta)\, F_1(u)- \frac{1+u}{u}\, F_2(u)\,,
\\
&g_1(u)= \frac{(1+u)(d-\Delta)}{u(2+u)}\,F_1(u) - \frac{d+(1+u)^2}{u^2(2+u)}\,F_2(u)\,,
\end{align}
where
\begin{align}
&F_1(u)=  {\cal N} \left( 2u \right)^{-\Delta}\,_2F_1\!\left( \Delta,\frac{1-d+2\Delta}{2},1-d+2\Delta, -\frac{2}{u}\right), \nonumber
\\
&F_2(u)=  {\cal N} \left( 2u \right)^{-\Delta}\,_2F_1\!\left( \Delta+1,\frac{1-d+2\Delta}{2},1-d+2\Delta, -\frac{2}{u}\right) ,
\end{align}
with
\be
{\cal N}= \frac{\Gamma\!\left(\Delta+1 \right)}{2  \pi^{d/2}(d-1-\Delta)(\Delta-1) \,\Gamma\!\left(\Delta+1-\frac{d}{2}\right) }\,.
\ee

\subsubsection{Spin 2}
To make contact with previous results in the literature we will compare (\ref{eq:propagatorJgk}) for $J=2$ with
the result for massive symmetric spin 2 field in \cite{Naqvi:1999va}.
%
The solution of the equations of motion for a symmetric spin two propagator can be organized in five structures (including the trace part for now)
\begin{align}
G_{\mu_1 \mu_2;\nu_1 \nu_2}(u)=\sum_{i=1}^5 
A^{(i)}(u)\,
T^{(i)}_{\mu_1\mu_2;\nu_1\nu_2}\,,
\label{eq:NaqviResult}
\end{align}
where $T^{(i)}_{\mu_1\mu_2;\nu_1\nu_2}$ are the five independent structures
\begin{align}
&T^{(1)}_{\mu_1\mu_2;\nu_1\nu_2}=
g^{\mu_1\mu_2}g^{\nu_1\nu_2}\,,\\
&T^{(2)}_{\mu_1\mu_2;\nu_1\nu_2}=
\partial_{\mu_1}u\,\partial_{\mu_2}u\,\partial_{\nu_1}u\,\partial_{\nu_2}u \,,\\
&T^{(3)}_{\mu_1\mu_2;\nu_1\nu_2}=
\partial_{\mu_1}\partial_{\nu_1}u\,\partial_{\mu_2}\partial_{\nu_2}u +\partial_{\mu_1}\partial_{\nu_2}u\,\partial_{\mu_2}\partial_{\nu_1}u\,,\\
&T^{(4)}_{\mu_1\mu_2;\nu_1\nu_2}=
\partial_{\nu_1}u\,\partial_{\nu_2}u\,g^{\mu_1\mu_2} +\partial_{\mu_1}u\,\partial_{\mu_2}u\,g^{\nu_1\nu_2}\,,\\
&T^{(5)}_{\mu_1\mu_2;\nu_1\nu_2}=
\partial_{\mu_1}\partial_{\nu_1}u\,\partial_{\mu_2}u\,\partial_{\nu_2}u+ \partial_{\mu_2}\partial_{\nu_1}u\,\partial_{\mu_1}u\,\partial_{\nu_2}u
+ (\nu_1\leftrightarrow \nu_2) \,,
\end{align} 
and the specific form of the functions $A^{(i)}(u)$ is given in \cite{Naqvi:1999va}. However, 
there are only three symmetric and traceless structures that can be constructed from $T^{(i)}_{\mu_1\mu_2;\nu_1\nu_2}$. These correspond to following structures in the embedding formalism
\begin{align}
(W_{12})^2&\rightarrow  
\frac{T^{(3)}_{\mu_1\mu_2;\nu_1\nu_2}}{2} -\frac{T^{(1)}_{\mu_1\mu_2;\nu_1\nu_2}\big(1+d-u(2+u)\big)}{(1+d)^2}-\frac{T^{(4)}_{\mu_1\mu_2;\nu_1\nu_2}}{1+d}\,,\\
W_{12} (W_1\cdot X_2 )(W_2 \cdot X_1)&\rightarrow 
-\frac{u(1+u)(u+2)T^{(1)}_{\mu_1\mu_2;\nu_1\nu_2}}{(1+d)^2}+\frac{(1+u)T^{(4)}_{\mu_1\mu_2;\nu_1\nu_2}}{1+d}-\frac{T^{(5)}_{\mu_1\mu_2;\nu_1\nu_2}}{4}\,,\\
(W_1\cdot X_2)( W_2 \cdot X_1)^2&\rightarrow 
\frac{u^2(2+u)^2T^{(1)}_{\mu_1\mu_2;\nu_1\nu_2}}{(1+d)^2}+T^{(2)}_{\mu_1\mu_2;\nu_1\nu_2}-\frac{u(u+2)T^{(4)}_{\mu_1\mu_2;\nu_1\nu_2}}{1+d}\,.
\end{align}
Therefore, the functions in the expansion (\ref{eq:NaqviResult}) are given by
\begin{align}
A^{(2)}(u)&=g_2(u)\,,\ \ \ \ \ \ \ \ 
A^{(3)}(u)=\frac{1}{2}\,g_0(u)\,,\ \ \ \ \ \ \ \ \ 
A^{(5)}(u)=-\frac{1}{4}\,g_1(u)\,,\\
A^{(1)}(u)&=-\frac{ 1+d-u(2+u) }{(1+d)^2}\,g_0(u)
-\frac{u(1+u)(u+2) }{(1+d)^2}\,g_1(u)
+ \frac{u^2(2+u)^2 }{(1+d)^2}\,g_2(u)\,,\\
A^{(4)}(u)&=  -\frac{1}{1+d}\,g_0(u) 
+\frac{(1+u) }{1+d}\,g_1(u)
-\frac{u(u+2) }{1+d}\,g_2(u)\,.
\end{align}
Using the results of the previous section for $J=2$, we recover the results of \cite{Naqvi:1999va}.
This is the full result for the propagator up to contact terms. In section \ref{sec:SplitSpin2} we shall discuss in detail the contact terms for the spin 2 case.

%

\subsection{Bulk-to-boundary propagator}

In the  embedding formalism, the bulk-to-boudary propagator of a  spin  $J$ and dimension $\Delta$ field  has the simple form
\begin{equation}
\Pi_{\Delta,J}(X,P;W,Z)=\mathcal{C}_{\Delta,J}\,\frac{\big((-2P\cdot X)(W\cdot Z)+2(W\cdot P)(Z\cdot X)\big)^{J}}{(-2P\cdot X)^{\Delta+J}}\,.
\label{eq:eq propagador boundary}
\end{equation}
This is the unique structure compatible with conformal symmetry, which in this formalism is encoded by the constraint 
\begin{equation}
\Pi_{\Delta,J}(X,\lambda P;\alpha_{1}W,\alpha_{2}Z+\beta P)=\lambda^{-\Delta}(\alpha_{1}\alpha_{2})^{J}\Pi_{\Delta,J}(X,P;W,Z)\,,
\end{equation}
for arbitrary constants $\lambda$, $\alpha_1$, $\alpha_2$ and $\beta$. 
The normalization constant $\mathcal{C}_{\Delta,J}$ is fixed by considering the bulk-to-bulk propagator, properly normalised by
its short distance behaviour, and then sending one of the bulk points to the boundary, according to 
\begin{equation}
\lim_{\lambda\rightarrow\infty}\lambda^{\Delta}\Pi_{\Delta,J}(X,\lambda P+O(\lambda^{-1});W,Z)=
\Pi_{\Delta,J}(X,P;W,Z)\,.
\end{equation}
Let us check that this works for the bulk-to-bulk propagator computed in the previous section.
First we observe that the 
recurrence relation (\ref{eq:solution}) is simplified in the limit $u\rightarrow \infty$.
In this limit, this relation preserves the same asymptotic behaviour for all functions $h_k(u)$. Hence,
from the asymptotic behaviour of $h_0\left(u\right)$, we conclude that $h_k \approx s_k u^{-\Delta}$, and 
therefore  (\ref{eq:solution}) gives rise to a recursion relation  for $s_k$,
\begin{equation}
s_k=c_{k}\Big((d-2k+2J-1) (d-\Delta+J-2) \,s_{k-1}+(2-k+J)\,s_{k-2}\Big)\,.
\end{equation}
This equation has the following solution
\begin{equation}
s_k=s_0\,\frac{  J!}{k!(J-k)! }
\frac{(-1)^{k}  }{ 
\left(J+\Delta-k-1\right)_{k}}\,,
\end{equation}
which implies after the use of (\ref{eq:equacao entre g e f}) that
\begin{equation}
g_{k}(u)\approx s_0\,\frac{  J!}{k!(J-k)! }
\, \frac{ J+\Delta-1  }{\Delta-1}\,u^{-\Delta-k}\,.
\end{equation}
It is then clear that we 
recover the form of the bulk-to-boudary propagator (\ref{eq:eq propagador boundary}),  
\begin{equation}
\Pi_{\Delta,J}\!\left(X,P;W,Z\right)=
s_0\,2^\D\frac{\left(\Delta-1+J\right)}{\Delta-1}\frac{\big(2(W \cdot P)( X\cdot Z)- 2(W\cdot Z)(X\cdot P)\big)^{J}}{\left(-2X\!\cdot\!P\right)^{J+\Delta}}\,.
\end{equation}
The constant $s_0$ is fixed by the normalisation imposed by the delta function source in the propagator equation. We can just fix it by looking at the asymptotic behaviour of 
 the function $h_0(u)$, which fixes the normalisation constant  ${\cal C}_{\Delta,J}$ introduced in (\ref{eq:eq propagador boundary}) to be
\begin{equation}
{\cal C}_{\Delta,J}=\frac{ \left(J+\Delta-1\right) \Gamma(\Delta)}{2 \pi^{d/2}\left(\Delta-1\right)\Gamma(\Delta+1-h)}\,.
\label{eq:normalizationboundary}
\end{equation}
\subsection{Short distance limit}

Next we consider the short distance limit where $u\rightarrow 0$. Our goal is to check computations done in \cite{Giombi:2013yva}
that can also be done by directly computing the difference
between the short distance behaviour of spin $J$ propagators  of dimension $\Delta$  and $d-\Delta$. 
First, let us note that our solution for the spin $J$ propagator is based on the recursion relation (\ref{eq:RecursionRel}), where the seed is given by the scalar propagator (\ref{eq:h0}). It so happens that the scalar propagator diverges at short distances. However, 
the coefficients of all the divergent terms are invariant under $\Delta\rightarrow d-\Delta$. Therefore, since the 
recurrence relation is also invariant under this transformation, 
the difference of the spin $J$ propagators  of dimension $\Delta$  and $d-\Delta$ is finite in the limit  $u\rightarrow 0$.
Defining $\tilde{h}_0$ as the difference of $h_0$ for dimension $\Delta$  and $d-\Delta$ we can obtain, from the explicit result 
(\ref{eq:h0}), 
\be
\tilde{h}_0(u)=\sin\!\left(\frac{\pi}{2} ( d -2\D )\right)\sum_{k=0}^\infty \frac{\Gamma(d-\Delta+k)\,\Gamma(\Delta+k)}{\pi^{\frac{d+1}{2}}2^{d+k}k!\,\Gamma\!\left(\frac{1+d}{2}+k\right)}\,(-u)^k  \,.
\label{serieshtilde0}
\ee


To make contact with the computation of \cite{Giombi:2013yva} we only need to consider the difference of the  trace of the spin $J$ propagators.
In the embedding formalism the trace can be obtained simply by acting on the  propagator  with the operator
\be
\frac{1}{\left(J!(\frac{d-1}{2})_J \right)^2}\,(K_1\cdot K_2)^J\,,
\ee
where $K$ is defined in (\ref{Projector}) and we were careful with the numerical factor to obtain exactly the trace. 
We show in  appendix \ref{Ap:EmbeddingOperations} that in the limit $u\to 0$ the action of this operator on the difference of propagators $\tilde{\Pi}$ is 
\begin{equation}
(K_1\cdot K_2)^J \,\tilde{\Pi}_{\Delta,J}(X_{1},X_{2},W_{1},W_{2}) \approx 
(K_1\cdot K_2)^J(W_{12})^{J}\,\tilde{g}_{0}(u=0)\,.
\label{eq:propagatorJ}
\end{equation}
To compute $\tilde{g}_{0}(u=0)$ we need to use the  relation (\ref{eq:equacao entre g e f}) involving  a sum over all the 
$\tilde{h}_k(u=0)$, which in turn can be determined using the recursion relation (\ref{eq:solution}) and the series expansion (\ref{serieshtilde0}) of $\tilde{h}_0$.
We did this computation up to $J=12$ and verified that the result for the difference of the trace of the propagators exactly matches that of  \cite{Giombi:2013yva},
\be
g(J)\,\frac{\!(\Delta+J-1)(\Delta-J-d+1)\,\Gamma(\Delta-1)\,\Gamma(d-1-\Delta)\sin\!\left(\frac{\pi}{2} ( d -2\D )\right)}{2^d\pi^{\frac{d+1}{2}}\Gamma\!\left(\frac{1+d}{2}\right)}\,, 
\ee
where $g(J)$ is given by
\begin{align}
g(J)&=\frac{(2J+d-2)(J+d-3)!}{(d-2)!J!}\,, \ \ \ \ \ \ \ d\ge 3\,,
\\
g(0)&=1\,, \ \ \ \ \ \ g(J)=2\,, \ \ \ \ \ \ \  \ \ \  \ \ \  \ \ \  \ \ \  d=2 \,.
\end{align}



\subsection{Massless limit\label{sec:MasslessLimit}}

To analyse the massless limit of the spin $J$ bulk-to-bulk propagator let us introduce the new representation
\be
\Pi_{\Delta,J}=(W_{12})^{J}G(u)+W_1\cdot \nabla\left(\sum_{k=1}^{J}(W_{12})^{J-k}\left(W_2\cdot X_1\right)^k\left(W_1\cdot X_2\right)^{k-1}L_k(u)\right) .
\label{masslessrep}
\ee
Comparing with expression (\ref{eq:propagatorJgk}) we conclude that
\begin{align}
g_0(u)&=G(u)+L_1(u)\,,\\
g_k(u)&=-L_k'(u)+(k+1) L_{k+1}(u) \,,\ \ \ \ \ k=1,\dots,J-1\,,\\
g_J(u)&=-L'_J(u)\,.
\end{align}
These relations can be inverted to give the functions $L_k$ in terms of the functions $g_k$,
\be
L_{k}(u)=-\sum_{l=k}^{J} 
\frac{\Gamma(l+1)}{\Gamma(k+1)}
\underbrace{\int^u\dots\int du' }_{l-k+1}g_{l}(u') \,.
\ee
The function $G(u)$ follows after a simple manipulation, 
\begin{align}
G(u)&=g_0(u)+\sum_{l=1}^{J}\Gamma(l+1)\underbrace{\int^u \dots\int du'}_{l}g_l(u')\,,\nonumber\\ 
&=\sum_{l=0}^{J}\sum_{i=l}^{J}\left(-1\right)^{i+l}\frac{\Gamma^{2}(i+1)}{\Gamma(l+1)\,\Gamma(i-l+1)}\,h_i(u)=h_0(u)\,, 
\end{align}
where we used equation (\ref{eq:equacao entre g e f}) and $f_i^{(i+k)}(u)=h_i^{(k)}(u)$.

Expression (\ref{masslessrep}) for the bulk-to-bulk propagator is very convenient to study the massless limit.
In this case, gauge invariance implies that the propagator is always coupled to conserved   currents.
If the current is also traceless, 
then the functions $L_k$ in  (\ref{masslessrep}) do not contribute to physical processes (because their contribution vanishes after integrating by parts).
In other words, the function $G(u)$ is the only physical degree of freedom. 
This is the same result found in \cite{Mkrtchyan:2010pp} for every spin $J$. 
Notice that, one must be careful in intermediate calculations  because  the  gauge artifacts $L_k(u)$ diverge in the massless limit $\D \to J+d-2$ (this is clear from the explicit form of the  coefficient $a_k$ in (\ref{eq:ak})). In general, however, the conserved current is not traceless. The analysis of this case is more involved and was considered in \cite{Francia:2008hd}.
The main result is that the structures that couple to the multiple traces of the current also remain finite in the massless limit.
To clarify this point we now review   the spin 2 case.

\subsubsection{Graviton}

As explained above, it is important to isolate physical components from gauge artifacts. 
The massive spin 2 symmetric and traceless propagator (\ref{eq:NaqviResult}) can be rewritten in the form, 
\begin{align} 
G_{\mu_1 \mu_2 ;\nu_1 \nu_2} = \ &
\frac{\partial_{\mu_1}\partial_{\nu_1} u\,\partial_{\mu_2}\partial_{\nu_2} u+\partial_{\mu_1}\partial_{\nu_2} u\,\partial_{\mu_2}\partial_{\nu_1} u}{2} \,G(u) 
+g_{\mu_1\mu_2} g_{\nu_1 \nu_2} \,H(u)
 \nonumber\\
& +\partial_{(\mu_1} \big[\partial_{\mu_2)} \partial_{\nu_1} u \,\partial_{\nu_2}u \,X(u)\big]
+ \partial_{(\nu_1} \big[\partial_{\nu_2)} \partial_{\mu_1} u \, \partial_{\mu_2}u \,X(u)\big] 
\label{eq:PhyscalComponent} \\
&  +\partial_{(\mu_1} \big[\partial_{\mu_2)}u \,\partial_{\nu_1} u \,\partial_{\nu_2}u \,Y(u)\big] 
+\partial_{(\nu_1} \big[\partial_{\nu_2)}u \,\partial_{\mu_1} u \,\partial_{\mu_2}u \,Y(u)\big] \nonumber \\
& + \partial_{\mu_1}\big[\partial_{\mu_2}u\,Z(u)\big] \,g_{\nu_1 \nu_2} + \partial_{\nu_1} \big[\partial_{\nu_2}u \,Z(u)\big] \,g_{\mu_1\mu_2} \,,
\nonumber
\end{align}
where $(\,,)$ denotes symmetrization. Only the first line in this expression gives a finite contribution when coupled to a conserved symmetric tensor (not necessarily traceless). The physical components   $G(u)$ and $H(u)$ can be written in terms of $h_0(u)$ as 
\begin{align}
G(u) =&\ h_0(u)\,,\\
H(u)= &\,-\frac{1}{d\big(d-1+\D(\D-d)\big)}
\bigg(d \big(2d-4+\D(\D-d)  \big)\int_u^\infty du'\int_{u'}^\infty du'' h_0(u'')\nonumber\\
&\,-d(1+u)\int_u^\infty du' h_0(u')+ \big(d+\D(\D-d)\big)h_0\bigg)\,.
\label{eq:MasslessPhysicalComponetsGraviton}
\end{align}
Both functions  are regular in the massless limit and agree with  \cite{D'Hoker:1999jc}.




\section{Split representation of AdS propagators}
\label{SplitRep}
There is an alternative representation for bulk-to-bulk propagators which is often termed as split representation. The aim of this section is to introduce this representation for spin $J$ fields
and explicitly compute the propagator in some examples. We start by defining a basis of spin $J$ harmonic functions, denoted as $\Omega_{\nu,J}$. As will be shown, the propagator can be written as a linear combination of these functions.

\subsection{Spin $J$ harmonic functions in AdS}

\begin{figure}[t!]
\begin{centering}
\includegraphics[scale=0.55]{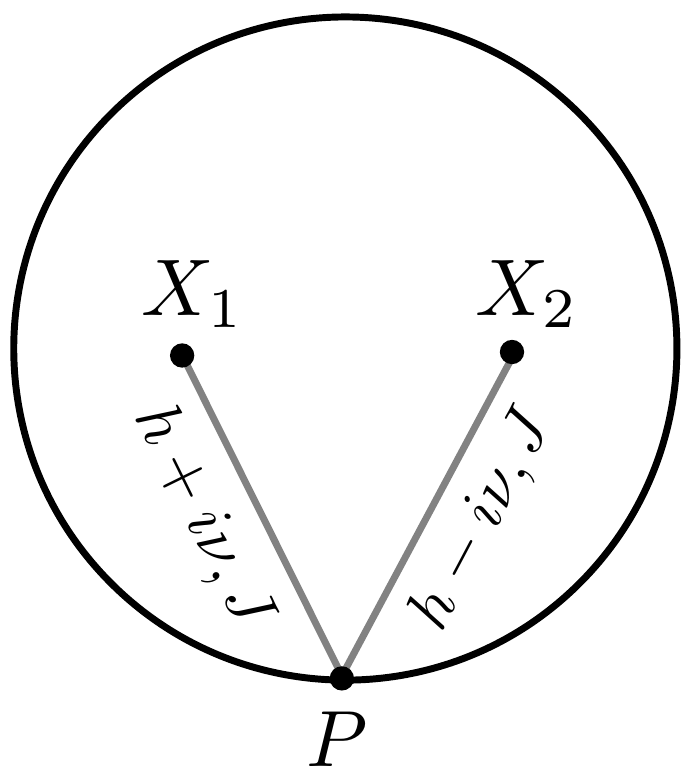}
\par\end{centering}
\caption{\label{fig:Omega}
Representation of AdS harmonic function $\Omega_{\nu,J}$ in terms of two spin $J$ bulk-to-boundary propagators of dimension $h\pm i\nu$
integrated over the boundary point.}
\end{figure}

The integral over the boundary point of the product of two bulk-to-boudary propagators, with dimensions $h+i\nu$ and $h-i\nu$, is by construction invariant under the exchange $\nu\leftrightarrow -\nu$. Moreover, it depends just on the bulk points $X_1$ and $X_2$ and polarization 
vectors $W_1$ and $W_2$. 
This is schematically represented in figure  \ref{fig:Omega} and leads to the following definition of the AdS harmonic function 
\begin{align}
\!\!\!
\Omega_{\nu,J}(X_1,X_2;W_1,W_2)
\!=\!
 \frac{\nu^2}{\pi  J!(h-1)_J}
 \!
\int_{\partial} \!dP \,\Pi_{h+i\nu,J}(X_1,P;W_1,D_Z)\Pi_{h-i\nu,J}(X_2,P;W_2,Z)\,,
\label{eq:OmegaIntegralBoundary}
\end{align} 
where $Z$ denotes a polarization vector on the boundary and $h=d/2$ was introduced for convenience. The operator $D_Z$, given by, 
\begin{align}
D_Z^{A}=\left(h-1+Z\cdot\frac{\partial}{\partial Z}\right)\frac{\partial}{\partial Z_{A}}-\frac{1}{2} Z^{A}\frac{\partial^2}{\partial Z\cdot \partial Z}\,,
\label{TodorovOper}
\end{align}
is the boundary counterpart of (\ref{Projector}) and implements index contraction of tensors defined on the boundary of AdS.
In appendix \ref{Ap:BulkToBoundary}, we compute   the integral (\ref{eq:OmegaIntegralBoundary}) and show that it is given by  a linear combination of two bulk-to-bulk propagators \cite{Leonhardt:2003qu}
\begin{align}
&\Omega_{\nu,J}(X_1,X_2;W_1,W_2)=\frac{i\nu}{2\pi} \Big(
\Pi_{h+i\nu,J}(X_1,X_2;W_1,W_2)-\Pi_{h-i\nu,J}(X_1,X_2;W_1,W_2) \Big)\,.
\label{integrationboundary}
\end{align}   
Notice that  $\Omega_{\nu,J}$  is an eigenfunction of the Laplacian operator and is divergence free, 
\begin{align}
\left(\nabla_{1}^{2}+h^2+\nu^2+J\right)\Omega_{\nu,J}(X_1,X_2;W_1,W_2) & =0\,,
\label{eq:LaplacianOmega}\\
\nabla_{1}\cdot K_{1}\,\Omega_{\nu,J}(X_1,X_2;W_1,W_2) & =0	\,.
\label{eq:divergenceOmega}
\end{align}
These properties follow from (\ref{integrationboundary}) and  equations (\ref{eq:Laplacian}-\ref{eq:divergence}) for the bulk-to-bulk propagator.

Besides being an eigenfunction of the Laplacian,  $\Omega_{\nu,J}$ satisfies an orthogonality relation.
To see this, consider the integral
\begin{align}
&
\frac{1}{J!\left(\frac{d-1}{2}\right)_J}
 \int_{AdS} dY\, \Omega_{\bar{\nu},J}(X_1,Y;W_1,K)\, \Omega_{\nu,J}(Y,X_2;W,W_2) 
 =
C_{\nu,\,\bar{\nu}}(X_1,X_2;W_1,W_2)\,, 
 \nonumber
\end{align}
This object can only depend on the invariants $X_1\cdot X_2$, $W_1\cdot W_2$ and $(W_1\cdot X_2)( W_2\cdot X_1)$. Therefore, it is invariant under the exchange 
$(X_1,W_1)\leftrightarrow (X_2,W_2)$.
By construction,   $C_{\nu,\,\bar{\nu}}$ is  an eigenfunction of the Laplacian. 
Thus, the expression
\begin{align}
\left(\nabla_1^2-\nabla_2^2\right)C_{\nu,\,\bar{\nu}}(X_1,X_2;W_1,W_2)=\left(\nu^2-\bar{\nu}^2\right)C_{\nu,\,\bar{\nu}}(X_1,X_2;W_1,W_2) =0\,,
\end{align}
must vanish because it must be both antisymmetric and  symmetric under the permutation
$(X_1,W_1)\leftrightarrow (X_2,W_2)$.
This means $C_{\nu,\,\bar{\nu}}$ only has support at $\bar{\nu}=\pm\nu$. Notice also that $C_{\nu,\bar{\nu}}$ is a harmonic function in the variables $X_1$ and $X_2$, so it should be proportional to $\Omega_{\nu,J}$. Thus
\begin{align}
C_{\nu,\,\bar{\nu}}(X_1,X_2;W_1,W_2)= \frac{1}{2}\big[\delta(\nu+\bar{\nu})+\delta(\nu-\bar{\nu})\big]\,\Omega_{\nu,J}(X_1,X_2;W_1,W_2)\,,
\label{eq:ConstraintFinal}  
\end{align}
where the constant of proportionality was determined in appendix \ref{Ap:orthoOmega}.
Integrating  $C_{\nu,\,\bar{\nu}}$ over $\nu$, we find
\be
\frac{1}{J!\left(\frac{d-1}{2}\right)_J}
 \int_{AdS} 
 \!\!\!
 dY\, \Omega_{\bar{\nu},J}(X_1,Y;W_1,K)
 \!\int\!
 d\nu \,\Omega_{\nu,J}(Y,X_2;W,W_2) 
 =
 \Omega_{\bar{\nu},J}(X_1,X_2;W_1,W_2)\,, 
\ee
which implies that
\begin{align}
\! \!\!\!
\int_{-\infty}^{\infty}
\! \!\!
d\nu\,\Omega_{\nu,J}(X_1,X_2;W_1,W_2)
\!=\!
\delta(X_1,X_2)(W_{12})^J
\!+\!
(W_1\cdot\nabla_1) (W_2\cdot\nabla_2)  Q  (X_1,X_2;W_1,W_2 )\,,
\label{eq:Completeness1}
\end{align}
 since we can always add to the right hand side of (\ref{eq:Completeness1}) a  total derivative because $\Omega_{\bar{\nu},J}$ has zero  divergence. The function $Q$ can be written as
\begin{align}
\!\!
Q (X_1,X_2;W_1,W_2)=
-\!\sum_{l=1}^J
\int \!
d\nu \,c_{J,l}(\nu)\big( (W_1\cdot\nabla_1)(W_2\cdot\nabla_2)\big)^{l-1}\Omega_{\nu,J-l}(X_1,X_2;W_1,W_2)\,.
\label{eq:Completeness1cCoefficients}
\end{align}
This means that we can write the completeness relation
\footnote{We call this a completeness relation because if we use the representation (\ref{eq:OmegaIntegralBoundary}) of the harmonic functions,
equation (\ref{eq:CompletenessNice})   tells us that the functions $F_{\nu,l,P}(X,W)\equiv(W\cdot \nabla)^l\Pi_{h+i\nu,J-l}(X,W;P,Z)$,
with $\nu \in \mathbb{R}$, $P\in \mathbb{R}^d$  and $l=0,1,\dots  J$, form a complete basis for spin $J$ (symmetric and traceless) tensors in AdS. 
}
\be
\sum_{l=0}^J\int d\nu \,c_{J,l}(\nu)\big( (W_1\cdot\nabla_1)(W_2\cdot\nabla_2)\big)^{l}\Omega_{\nu,J-l}(X_1,X_2;W_1,W_2) = 
\delta(X_1,X_2) (W_{12})^J\,,
\label{eq:CompletenessNice}
\ee
where $c_{J,0}(\nu)=1$. In  appendix \ref{Ap:orthoOmega} we  derive the general formula
\be
c_{J,l}(\nu)=
\frac{2^l (J-l+1)_l
   \left(h+J-l-\frac{1}{2}\right)_l}
   {l! (2 h+2 J-2 l-1)_l (h+J-l-i
   \nu )_l (h+J-l+i \nu )_l}\,,
   \label{cformula}
\ee
using a recursive argument to increase $J$ and $l$.

\subsection{Split representation
\label{subsec:split}}


\begin{figure}[t!]
\begin{centering}
\includegraphics[scale=0.55]{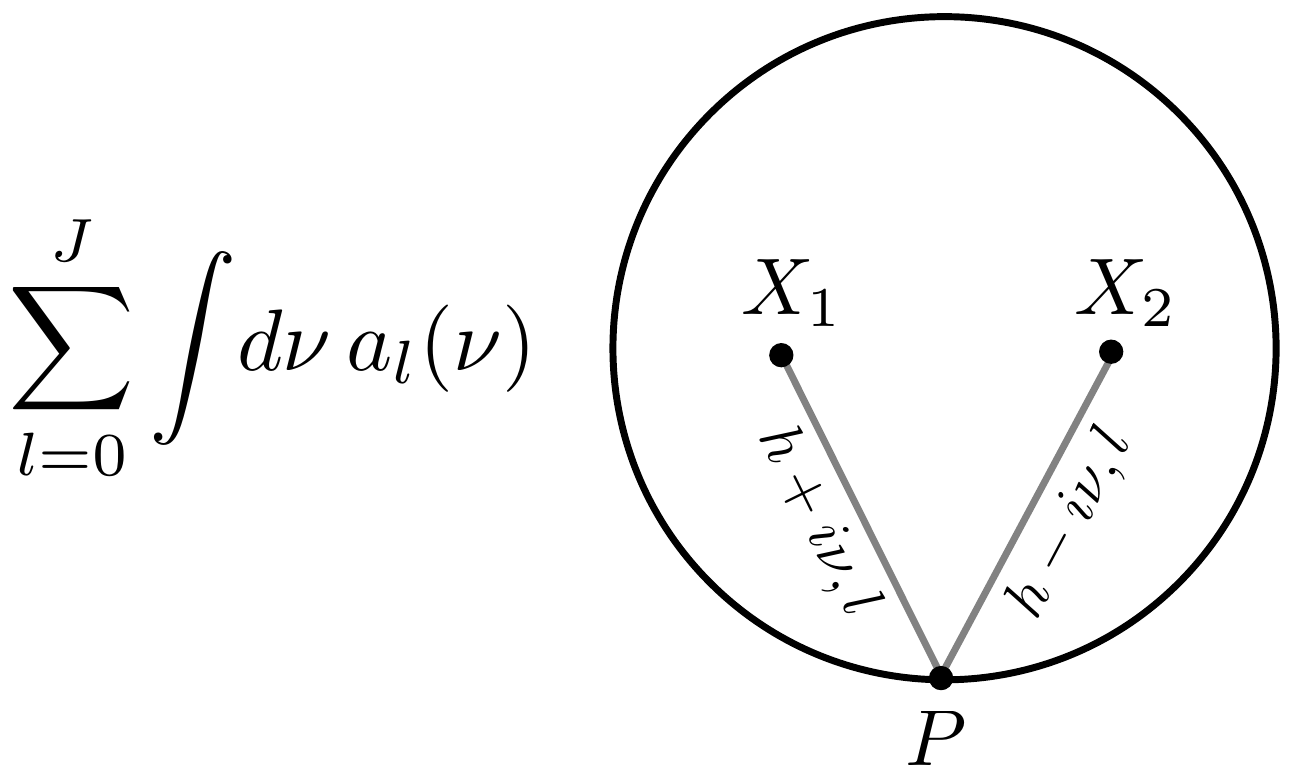}
\par\end{centering}
\caption{\label{fig:Propagator}
The split representation of the spin $J$ propagator 
obtained by integrating over $\nu$ and summing over the spin $l$ of 
two bulk-to-boundary propagators of dimension $h\pm i\nu$ integrated 
over the boundary point, according to (\ref{eq:SlipRepStart}).}
\end{figure}

Let us express  the propagator as a linear combination of harmonic functions,
\begin{align}
\Pi_{\Delta,J}(X_1,X_2;W_1,W_2)&=\sum_{l=0}^J\int d\nu \,a_{l}(\nu)\big((W_{1}\cdot\nabla_{1})( W_{2}\cdot \nabla_{2})\big)^{J-l} \, \Omega_{\nu,l}(X_1,X_2;W_1,W_2)\,,
\label{eq:SlipRepStart}
\end{align}
as represented in figure  \ref{fig:Propagator}. This is the split representation of the propagator.
The coefficients $a_l(\nu)$ of the split representation can be obtained using the equations of motion. More specifically, using the commutation relation
\be
\left[\nabla^2,(W\cdot\nabla )^n\right]=-n (2h-1+2W\cdot \partial_W-n)(W\cdot \nabla )^n\ ,
\ee 
in equation (\ref{eq:Laplacian}), we find 
\begin{align}
&
\big[\D(\D-d)-J-\nabla_1^2\big] \, \Pi_{\D,J}(X_1,X_2;W_1,W_2)=
\label{eq:propogatorEq}
\\
&
\sum_{l=0}^J\!\int\! d\nu \,a_l(\nu)\big[(J-l)(2h+J+l-2)+\nu^2+(\D-h)^2\big]\big( (W_1\cdot\nabla_1 )(W_2\cdot\nabla_2)\big)^{J-l}\Omega_{\nu,l}\,.\nonumber
\end{align}
 We shall assume that  the traceless part of the contact terms has the following general form
\begin{align}
 &\big[\D(\D-d)-J-\nabla_1^2\big]\,\Pi_{\D,J}(X_1,X_2;W_1,W_2) =
\label{eq:contact_terms_structure}
\\& 
\sum_{s=0}^J v_{J-s}  \big( (W_1\cdot\nabla_1)(W_2\cdot\nabla_2)\big)^{J-s}\big[ (W_{12})^s \d(X_1,X_2)\big]\,, 
\nonumber
\end{align}
where $v_{J-s}$ are constants and $v_0=1$ is fixed by the normalisation of the delta-function term without derivatives.
Using the representation    (\ref{eq:CompletenessNice}) of the delta function, and comparing with (\ref{eq:propogatorEq}) we obtain
\be
a_l(\nu)\big[(J-l)(2h+J+l-2)+\nu^2+(\D-h)^2\big]=
\sum_{s=l}^J v_{J-s} \,c_{s,s-l}(\nu)\,,
\label{eq:alnu}
\ee
for $0\le l\le J$. Since $c_{J,0}=1$, the case $l=J$ gives immediately 
\begin{align}
a_{J}(\nu)=\frac{1}{\nu^2+(\D-h)^2}\,.
\label{eq:splita1}
\end{align}
To determine the  coefficients $a_l(\nu)$ for $l<J$ 
we look for a solution of
(\ref{eq:alnu}) with the minimal number of poles in $\nu$. From the explicit form of $c_{s,s-l}(\nu)$ given in  (\ref{cformula}), we are led to the ansatz
\begin{align}
a_l(\nu)=\sum_{q=1}^{J-l}\frac{r_{l,q}}{\nu^2+(h+l+q-1)^2}\,,\ \ \ \ \ \ \ \ l<J\,,
\label{eq:alGenericForm}
\end{align}
where the $J$ and $\D$ dependence of the residue is implicit.
To determine these residues we consider equation (\ref{eq:alnu}) when $\nu^2 \to -(h+l+q-1)^2$. This gives
\begin{align}
&r_{l,q}\big[(J-l)(2h+J+l-2)-(h+l+q-1)^2+(\D-h)^2\big]=
\nonumber
\\
&\sum_{s=l}^J v_{J-s} \,\lim_{\nu\to i(h+l+q-1)}
\left[\nu^2+(h+l+q-1)^2\right]c_{s,s-l}(\nu)=
\label{eq:rintermsofv}\\
&\sum_{s=l+q}^J v_{J-s} 
\frac{(-1)^{q+1} 2^{s-l+1} s!
   (h+l+q-1)  
   \left(h+l-\frac{1}{2}\right)_{s-l
   }}{(q-1)! l!(s-l)! (s-l-q)! (2 h+2
   l-1)_{s-l} (2 h+2 l+q-1)_{s-l}}\,.
 \nonumber
\end{align}
On the other hand, the limit $\nu^2 \to \infty$ of  (\ref{eq:alnu}) gives
\be
v_{J-l} =\sum_{q=1}^{J-l} r_{l,q} \,.
\ee
Using these two equations, one finds that
\be
v_k=
\frac{\left(-1\right)^k J!
   (2 h+2 J-2 k-2)_k}{2^k k! (J-k)!
   (h+J-k-1)_k (\Delta+J-k -1)_k (2
   h-\Delta+J-k -1)_k}\,,
   \label{eq:explicitv}
\ee
which in turn determines the residues $r_{l,q}$ through equation (\ref{eq:rintermsofv}).
As an example, we show the first   coefficients,
\begin{align} 
\!\!
a_{J-1}(\nu)&=-\frac{J}{y_2\left(\nu^2+\left(h+J-1\right)^2\right)}\,,
\label{eq:splita2}\\ 
\!\!
a_{J-2}(\nu)&=\frac{J\left(J-1\right)}{4y_3\left(J+h-2\right)\!\left(\nu^2+\left(h+J-1\right)^2\right)}-
\frac{J\left(J-1\right)}{4y_2\left(J+h-2\right)\!\left(\nu^2+\left(h+J-2\right)^2\right)}\,,
\label{eq:splita3}
\end{align} 
with $y_k=(\Delta+J-k)(2h-\Delta+J-k)$. The expression for other coefficients,  $a_{J-k}(\nu)$,  cannot be written explicitly in such a compact form. However, they are completely determined by
(\ref{eq:rintermsofv}) and  (\ref{eq:explicitv}).
In  section \ref{CPWdecomposition}, we shall use a  very different argument to derive a recursion relation that also fixes all coefficients $a_l(\nu)$.

In finding the solution for the coefficients $a_l(\nu)$ we made two simplifying assumptions: the structure of the contact terms
in the propagator equation (\ref{eq:contact_terms_structure}) and the existence of a minimal number of poles in $\nu$.
In the next subsections, we consider the case of spin 1 and 2 and determine the full split representation.
We will find that the above assumptions are indeed correct.

\subsection{Spin 1}

The divergence of the spin 1 propagator vanishes at separate points. However, one must take special care with possible contact terms. Notice that, in general,  these give non-zero contributions to Witten diagrams.
To determine the possible contact terms we return to the Proca equation, which gives for the propagator
\begin{align}
 \left(D^2_1 +1-\D(\D-d)\right)\Pi_{\mu ,\nu }(x_1,x_2) 
 -  D_{\mu } D^\sigma \Pi_{ \sigma , \nu  } (x_1,x_2)  =
 - g_{\mu  \nu } 
\delta(x_1,x_2) \,.
\end{align}
In the embedding formalism this can be written as
\be
\left[\nabla^2_1 +1-\D(\D-d)- \frac{2}{d-1} (W_1\cdot \nabla_1)( K_1\cdot \nabla_1)\right]
 \Pi_{\Delta,1} (X_1,X_2;W_1,W_2 )   
= - W_{12}\,
\delta(X_1,X_2)\,. 
\nonumber
\ee
It is straightforward to check that this equation is solved exactly by the split representation we found in the previous section,
\begin{align}
\Pi_{\Delta,1}(X_1,X_2;W_1,W_2)=& \int
\frac{ d\nu \, \Omega_{\nu,1}(X_1,X_2;W_1,W_2)}{ \nu^2+ (\D-h)^2 }
\label{eq:SlipRepStart1} \\
&-\int 
\frac{d\nu \, (W_{1}\cdot\nabla_{1})( W_{2}\cdot \nabla_{2})  \, \Omega_{\nu,0}
(X_1,X_2 )}{(\D-1)(2h-\D-1)\left(\nu^2+ h^2\right)}\,.
\nonumber
\end{align}
In this calculation we used the basic properties 
(\ref{eq:LaplacianOmega}) and (\ref{eq:divergenceOmega}) of the harmonic functions, the simple commutators (\ref{eq:BasicCommutators}) 
given in appendix \ref{Ap:orthoOmega} and the completeness relation (\ref{eq:CompletenessNice}).
We conclude that, in this case, the coefficients $a_1(\nu)$ and $a_0(\nu)$ are entirely determined by their poles (\ref{eq:splita1}) and (\ref{eq:splita2}) without any additional regular piece.

\subsection{Spin 2}
\label{sec:SplitSpin2}

The massive spin 2 propagator is traceless and divergenceless when the two bulk points it connects are different.
To determine possible contact terms we write the full propagator as a sum of three terms. The first term is the traceless part that we discussed so far.
Since it is traceless it can be written as a polynomial in $W_1$ and $W_2$,
\begin{align}
\Pi_{\Delta,2}(X_1,X_2;W_1,W_2)&=\sum_{l=0}^2\int d\nu \,a_{l}(\nu)\big((W_{1}\cdot\nabla_{1})( W_{2}\cdot \nabla_{2})\big)^{2-l} \, \Omega_{\nu,l}(X_1,X_2;W_1,W_2)\,, 
\label{eq:SlipRepStart2}
\end{align}
where the poles of the coefficients were computed in  section
\ref{subsec:split},
\begin{align}
a_{2}(\nu)&= \frac{1}{ \nu^2+ (\D-h)^2 }\,,\ \ \ \ \ \ \ \ \ \ \ \ 
a_{1}(\nu)=-\frac{2}{\D(2h-\D)\left(\nu^2+ (h+1)^2\right)}\,,
\\
a_{0}(\nu)&=\frac{1}{2h(\D-1)(2h-\D-1)  
\left(\nu^2+ (h+1 )^2\right)}-
\frac{1}{2h\D(2h-\D) \left(\nu^2+h^2\right)}\,.
\end{align}
We will show that this is the complete expression for these coefficients. 
In addition to the traceless part of the propagator, we add a second term which is a pure trace,
\begin{equation}
 \left(\eta^{AB}+X_1^AX_1^B\right)\left(\eta^{CD}+X_2^C X_2^D\right) \int d\nu\, t(\nu)\, \Omega_{\nu,0} (X_1,X_2)\,,
 \label{abar1}
\end{equation}   
and a third term given by
\begin{equation}
 \Big[\left(\eta^{AB}+X_1^AX_1^B\right) \nabla_2^C \nabla_2^D +\nabla_1^A\nabla_1^B 
\left(\eta^{CD}+X_2^C X_2^D\right) \Big] \int d\nu \,q(\nu)\, \Omega_{\nu,0}  (X_1,X_2)\,.
 \label{abar2}
\end{equation}  

We now use the full equation of motion (\ref{eomspin2massive})  to write the complete equation for the propagator
\begin{align}
\!\! &\big(D^2_1 +2-\D(\D-d)\big)G_{\mu_1 \mu_2;\nu_1 \nu_2}
 \!-\!  
 D_{\mu_1} D^\sigma G_{ \sigma \mu_2; \nu_1 \nu_2 }
 \!-\!  
 D_{\mu_2} D^\sigma G_{\mu_1 \sigma ; \nu_1 \nu_2}
  \!-\!  
2g_{\mu_1 \mu_2} G_{\sigma}{}^ \sigma{}_{;\nu_1 \nu_2}+
\label{eq:SpinTwoEquationsOfMotion}\\
\!\! &\bigg(D_{\mu_1} D_{\mu_2}-\frac{\D(\D-d)}{d-1} 
g_{\mu_1 \mu_2}\bigg)G_{\sigma}{}^ \sigma{}_{;\nu_1 \nu_2} =
 -\frac{1}{2}\Bigl(g_{\mu_1 \nu_1}g_{\mu_2 \nu_2} +g_{\mu_1 \nu_2}g_{\mu_2 \nu_1} -  {2g_{\mu_1\mu_2}g_{\nu_1 \nu_2}\over d-1}\Bigr)
\delta(x_1,x_2)\,.
\nonumber
\end{align}
We can determine the coefficients 
$t(\nu)$ and $q(\nu)$ by
imposing this equation,
including contact terms. 
To see this, let us apply the left hand side
to each one of the   terms
(\ref{eq:SlipRepStart2}), (\ref{abar1}) and (\ref{abar2}), that make up the full propagator.
Acting on the traceless part of the propagator, the left hand side of (\ref{eq:SpinTwoEquationsOfMotion}) gives
a traceless contribution 
\be
-(W_{12})^2 \d(X_1,X_2)+
\frac{(2h-1)^2 \big((W_1 \cdot \nabla_1)(W_2 \cdot \nabla_2)\big)^2}{2h(2h+1)\D(\D-1)(2h-\D)(2h-\D-1)}\,\d(X_1,X_2)\,,
\ee
plus a contribution with non zero trace,
\be
\int d\nu \,\rho(\nu) \left( \eta^{AB}+X_1^AX_1^B\right) \left[
\nabla_2^C\nabla_2^D+
\frac{\nu^2+h^2}{2h+1}
\left( \eta^{CD}+X_2^CX_2^D\right)
\right] \Omega_{\nu,0}(X_1,X_2)\,,
\ee
where
\be
\rho(\nu)=\frac{2}{(2h+1)\D(2h-\D)}-\frac{
2(2h-1) (\nu^2+h^2)
}{(2h+1)^2\D(\D-1)(2h-\D)(2h-\D-1)}\,.
\ee
Applying the left hand side of (\ref{eq:SpinTwoEquationsOfMotion}) to (\ref{abar1}), we obtain
\begin{equation}
\int d\nu \,t(\nu)\Big[ \sigma(\nu)\left(\eta^{AB}+X_1^AX_1^B\right) + (2h-1)\nabla_1^A\nabla_1^B \Big] 
\left(\eta^{CD}+X_2^C X_2^D\right)\Omega_{\nu,0} (X_1,X_2)\,,
\end{equation}
where
\be
\sigma(\nu)=\frac{4h(\D-1)(2h-\D-1)}{2h-1}-\nu^2-h^2\,.
\ee
Finally,  the left hand side of (\ref{eq:SpinTwoEquationsOfMotion}) applied to (\ref{abar2}) gives
\begin{align}
\int d\nu \,q(\nu) &\bigg(
\Big[ \sigma(\nu)\left(\eta^{AB}+X_1^AX_1^B\right) + (2h-1)\nabla_1^A\nabla_1^B \Big]\nabla_2^C \nabla_2^D+
\\
& \D(2h-\D)\left[ 
\nabla_1^A\nabla_1^B -\frac{\nu^2+h^2}{2h-1}\left(\eta^{AB}+X_1^AX_1^B\right)\right]\left(\eta^{CD}+X_2^C X_2^D\right)\bigg)
\Omega_{\nu,0} (X_1,X_2) \,.
\nonumber
\end{align}
To perform these calculations the following identities were useful
\begin{align}
&\nabla^2 \nabla^A\nabla^B F(X) =
\nabla^A\nabla^B \left[\nabla^2-2(2h+1)\right]F(X)+ 2 \left(\eta^{AB}+X^AX^B\right) \nabla^2 F(X)\,,
\nonumber
\\&
\nabla^A\nabla_C\nabla^C\nabla^B F(X)+
\nabla^B\nabla_C\nabla^A\nabla^C F(X) =
\nabla^A\nabla^B\left(2\nabla^2-4h\right)F(X)\,,
\end{align}
where $F(X)$ is a scalar function in AdS and the covariant derivative   was defined in (\ref{generalnabla}).

Putting together the contributions from the three terms and requiring that they sum up to the right hand side of (\ref{eq:SpinTwoEquationsOfMotion}), determines
\begin{align}
q(\nu)&=-\frac{
2h-1
}{2h(2h+1)\D(\D-1)(2h-\D)(2h-\D-1)}\,,
\nonumber
\\
t(\nu)&=q(\nu)\left[\frac{
\nu^2+h^2}{2h+1}-\frac{\D(2h-\D)}{2h-1}
\right]. 
\end{align}
Since both $t(\nu)$ and $q(\nu)$ are analytic in $\nu$, we conclude that the terms
(\ref{abar1}) and (\ref{abar2}) are pure contact terms, as expected.

It should be possible to generalize this analysis  to propagators with higher spin, using the appropriate equations of motion \cite{Singh:1974qz,Rindani:1985pi,Aragone:1988yx,Bouatta:2004kk,Francia:2007ee}. However, we will not attempt here to find a closed formula for the contact terms of a propagator of arbitrary   spin.

\subsubsection{Graviton propagator }

In the massless limit, the bulk propagator couples to a conserved current. This means one can drop total derivatives of the propagator. The full propagator is then given by
\begin{align}
&\int d\nu \,a_2(\nu) \,\Omega_{\nu,2}^{AB,CB}(X_1,X_2)
+({\rm total\ derivative})
\nonumber\\
&
+\left(\eta^{AB}+X_1^A X_1^B\right)\left(\eta^{CD}+X_2^C X_2^D\right) 
\int d\nu \left[t(\nu)+a_0(\nu)\,\frac{(h^2+\nu^2)^2}{(2h+1)^2} \right]\Omega_{\nu,0}(X_1,X_2) \,.
\label{splitminusder}
\end{align}
The graviton propagator is obtained in the massless limit $\D \to d=2h$. Recall that there is no van Dam-Veltman-Zhakarov discontinuity in AdS \cite{Kogan:2000uy,Porrati:2000cp}.
In this limit, both $t(\nu)$ and $a_0(\nu)$ diverge but the combination that appears in
(\ref{splitminusder}) remains finite. This gives 
the split representation of the graviton propagator
\begin{align}
\Pi_{d,2}^{AB,CD}(X_1,X_2)=&\int\frac{ d\nu }{\nu^2+h^2}\, \Omega_{\nu,2}^{AB,CD}(X_1,X_2)
+(total\ derivative)
  \label{SplitGraviton}
  \\
&
-\frac{\left(\eta^{AB}+X_1^A X_1^B\right)\left(\eta^{CD}+X_2^C X_2^D\right) }{2 h(2h -1)  }
\int \!\frac{d\nu}{
(h+1)^2+\nu ^2 } \, \Omega_{\nu,0}(X_1,X_2) \,.
\nonumber 
\end{align}
Notice that the total derivative (or pure gauge term) diverges in the massless limit.

\section{Three-point function}
\label{OPECubicAdScoupling}
\begin{figure}
\begin{centering}
\includegraphics[scale=0.55]{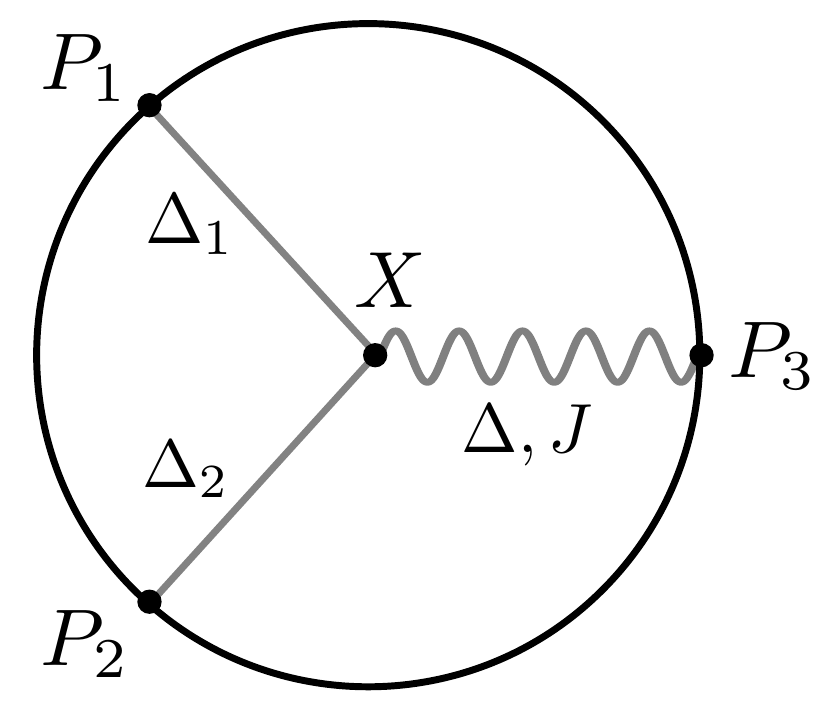}
\par\end{centering}
\caption{\label{fig:CubicAdS} Witten diagram that computes a CFT three-point function of a spin $J$ primary operator of dimension $\D$
and  two scalar primary operators of dimension $\D_1$ and $\D_2$.}
\end{figure}

A simple application of the above formalism is to consider a three-point Witten diagram with 
two insertions of scalar fields $\phi_1$, $\phi_2$ and one of a spin $J$ field at the boundary,
as represented in figure \ref{fig:CubicAdS}. 
This diagram, computed at tree level, will allow us to related the OPE coefficient
of the dual operators in the field theory to the local coupling of AdS fields.
 The simplest AdS local cubic vertex of a spin $J$ field to two scalars $\phi_1$ and $\phi_2$  is of the form
\be
 g_{\phi_1\phi_2h}  \int_{\rm AdS} d  x  \sqrt{g} \,    \big(\phi_2 \nabla_{\mu_1} \cdots \nabla_{\mu_J}  \phi_1\big) 
 h^{\mu_1\cdots \mu_J}\,, 
 \label{eq:bulkinteraction}
\ee
where $g_{\phi_1\phi_2h}$ is a bulk coupling 
 constant.  Notice that the derivatives can act on either of the scalar fields because we  consider a spin $J$ field of vanishing divergence.
Moreover,  a vertex with more derivatives can be reduced to this form by integrating by parts and using the (linear) equations of motion of the fields.

To compute the cubic Witten diagram in figure \ref{fig:CubicAdS} from the above vertex, we consider insertions of the scalar field $\phi_1$ at the boundary point $P_1$, of $\phi_2$ at  $P_2$
and of the spin $J$ field at $P_3$. The corresponding bulk-to-boudary propagator for a field of dimension 
$\Delta$ and spin $J$ is given in (\ref{eq:eq propagador boundary}). Thus,
this Witten diagram is given by the integral over the AdS interaction point of the bulk-to-boudary propagators, and generates
the CFT three-point function
\footnote{
\label{2ptnormalization}
The factor
$1/\sqrt{
\mathcal{C}_{\Delta_1}\mathcal{C}_{\Delta_2}\mathcal{C}_{\Delta,J}
}$ corresponds to the normalization choice of CFT operators that have unit two-point function,
$$
\left\langle {\cal O}(P_1,Z_1)  
{\cal O} (P_2,Z_2) \right\rangle=
\frac{\big( (-2P_1\cdot P_2)( Z_2\cdot Z_1) +2(Z_2\cdot P_1 )( Z_1\cdot P_2) \big)^J}{ (-2P_1\cdot P_2)^{\Delta+J}}
\,.
$$
} 
\begin{align}
&
\left\langle {\cal O}_{\phi_1}(P_1)\,{\cal O}_{\phi_2}(P_2)\,{\cal O}_{h}(P_3,Z) \right\rangle=
 \nonumber\\
& \frac{g_{\phi_1\phi_2h} }{\sqrt{
\mathcal{C}_{\Delta_1}\mathcal{C}_{\Delta_2}\mathcal{C}_{\Delta,J}}
}   \int_{\rm AdS} dX\,
\Pi_{\Delta_2,0} (X,P_2)
\frac{\Pi_{\Delta,J} (X,P_3;K,Z)
\left(W \cdot \nabla \right)^J 
\Pi_{\Delta_1,0} (X,P_1)}{
J!\left(\frac{d-1}{2}\right)_J}=
\label{cubic}  \\
 & \frac{g_{\phi_1\phi_2h} }{\sqrt{
\mathcal{C}_{\Delta_1}\mathcal{C}_{\Delta_2}\mathcal{C}_{\Delta,J}}
} 
  \,b(\Delta_1,\Delta_2,\Delta,J) \,
\frac{\big((Z\cdot P_1)P_{23} -(Z\cdot P_2 )P_{13}\big)^J}{
P_{12}^{\frac{\Delta_1+\Delta_2-\Delta+J}{2}}
P_{13}^{\frac{\Delta_1+\Delta-\Delta_2+J}{2}}
P_{23}^{\frac{\Delta+\Delta_2-\Delta_1+J}{2}}
}\,,
\nonumber
\end{align}
where $K$ is the projector defined in (\ref{Projector}) and we used the notation $P_{ij}=-2P_i\cdot P_j$ and $\mathcal{C}_{\Delta}=\mathcal{C}_{\Delta,0}$ for short.
In the last equality, we used the fact that this three-point function is determined by conformal symmetry up to an overall constant.
To determine the constant $b(\Delta_1,\Delta_2,\Delta,J) $ we have to perform the integral over AdS.
In the case  $J=0$, the AdS integral in (\ref{cubic})  is well known \cite{Rastelli3pt,ourBFKL} and gives 
\be
b(
\Delta_1,\Delta_2,\Delta,0)
\!=\!
\mathcal{C}_{\Delta_1}\mathcal{C}_{\Delta_2}\mathcal{C}_{\Delta}
 \frac{\pi^{\frac{d}{2}}
 \Gamma\!\left(\frac{\Delta_1+\Delta_2+\Delta-d}{2}\right)
\Gamma\!\left(\frac{\Delta_1+\Delta_2-\Delta}{2}\right)
\Gamma\!\left( \frac{\Delta_1+\Delta-\Delta_2}{2}\right)
 \Gamma\!\left(\frac{\Delta+\Delta_2-\Delta_1}{2}\right)}{
 2\, \Gamma(\Delta_1) \,\Gamma(\Delta_2)\, \Gamma(\Delta)} \,.
\ee
To compute the integral for general spin $J$, we use the differential operator
\be
D_{31}= Z\cdot P_1 \left( Z\cdot \frac{\partial}{\partial Z}-P_3\cdot\frac{\partial}{\partial P_3} \right) 
+ P_3\cdot P_1   \left( Z\cdot \frac{\partial}{\partial P_3}\right),
\label{eq:NewDifferentialOP}
\ee
introduced in \cite{Costa:2011dw}.
Noting that
\begin{align}
&\frac{1}{J!\left(\frac{d-1}{2}\right)_J} \,\Pi_{\Delta,J} (X,P_3;K,Z)
\left(W \cdot \nabla \right)^J \Pi_{\Delta_1,0} (X,P_1)
=\nonumber\\
&=2^J \mathcal{C}_{\Delta_1} \mathcal{C}_{\Delta,J} \, \frac{(\Delta_1)_J}{(\Delta)_J}
\,(D_{31})^J
\frac{1}{(-2P_3\cdot X)^{\Delta}(-2P_1\cdot X)^{\Delta_1+J}}
\,,
\label{eq:DerivativeTrick}
\end{align}
the computation of the above AdS integral reduces to that of the scalar case by commuting the differential operator $(D_{31})^J$ with the integral symbol.
Finally, using
\begin{align}
&
\frac{\big((Z\cdot P_1)  P_{23} -(Z\cdot P_2 ) P_{13}\big)^J}{
P_{12}^{\frac{\Delta_1+\Delta_2-\Delta+J}{2}}
P_{13}^{\frac{\Delta_1+\Delta-\Delta_2+J}{2}}
P_{23}^{\frac{\Delta+\Delta_2-\Delta_1+J}{2}}
}=
\nonumber
\\
&
=\frac{1}{ \left(\frac{\Delta+\Delta_2-\Delta_1-J}{2}\right)_J}
\,(D_{31})^J
\frac{1}{
P_{12}^{\frac{\Delta_1+\Delta_2-\Delta+J}{2}}
P_{13}^{\frac{\Delta_1+\Delta-\Delta_2-J}{2}}
P_{23}^{\frac{\Delta+\Delta_2-\Delta_1+J}{2}}
}\,,
\label{eq:DerivativeTrick3}
\end{align}
we arrive at the result
\begin{align}
 &b(\Delta_1,\Delta_2,\Delta,J)=2^J
 \left(\frac{\Delta+\Delta_2-\Delta_1-J}{2}\right)_{\!\!J}\,
\frac{ \mathcal{C}_{\Delta,J}}{ \mathcal{C}_{\Delta,0}}
\frac{ \mathcal{C}_{\Delta_1}}{ \mathcal{C}_{\Delta_1+J}}
 \frac{(\Delta_1)_J}{(\Delta)_J}
 \, b(\Delta_1+J,\Delta_2,\Delta,0)=
  \label{eq:bfunction}
 \\
 &=\mathcal{C}_{\Delta_1}\mathcal{C}_{\Delta_2}\mathcal{C}_{\Delta,J} 
   \,
 \frac{\pi^{\frac{d}{2}}
 \Gamma\!\left(\frac{\Delta_1+\Delta_2+\Delta-d+J}{2}\right)
 \Gamma\!\left(\frac{\Delta_1+\Delta_2-\Delta+J}{2}\right)
 \Gamma\!\left( \frac{\Delta+\Delta_1-\Delta_2+J}{2}\right)
  \Gamma\!\left(\frac{\Delta+\Delta_2-\Delta_1+J}{2}\right)}{
 2^{1-J}\, \Gamma(\Delta_1)\, \Gamma(\Delta_2)\, \Gamma(\Delta+J)} \,.
 \nonumber
\end{align}
The result  (\ref{cubic}) establishes the relation between the local AdS coupling  $g_{\phi_1\phi_2h}$ and the CFT  
OPE
coefficient $C_{\phi_1\phi_2h}$,
\be
C_{\phi_1\phi_2h} = 
 \frac{b(\Delta_1,\Delta_2,\Delta,J)}{\sqrt{\mathcal{C}_{\Delta_1}\mathcal{C}_{\Delta_2}\mathcal{C}_{\Delta,J}}}\,g_{\phi_1\phi_2h} \,.
 \label{Cfromg}
\ee

As a check of this result, let us consider the case of the stress-energy tensor with $\D=d$ and $J=2$.
In this case, the OPE coefficient is determined by a Ward identity \cite{OsbornCFTgeneraldim, Dolan:2000ut}
\be
C_{\phi\phi T_{\mu\nu}}  = 
 \frac{d \,\D_\phi}{(d-1) \sqrt{C_T}}\,,
 \label{WardT}
\ee
where $C_T$ is the coefficient of the two point function of the (standard) stress tensor  (notice that here we are redefining the stress tensor such that it has unit two point function). 
This is given by \cite{Liu:1998bu,Kovtun}
\be
C_T=\frac{1}{2\pi G_N } \frac{d+1}{d-1}
\frac{\pi^{\frac{d}{2}}\Gamma(d+1)}{\Gamma^3\!\left(\frac{d}{2}\right)}\,,
\label{CTfromGN}
\ee  
where $G_N $ is the gravitational coupling of the $(d+1)$-dimensional (Euclidean) dual theory
\be
S_E=\frac{1}{16\pi G_N } \int d^{d+1}x 
\sqrt{g}\, \left[-d(d-1)-\mathcal{R}+ \frac{1}{2}(\nabla \phi)^2 +  \frac{1}{2}M_\phi^2 \phi^2\right],
\ee
and we are setting the AdS radius to one.
Expanding this action around the AdS background,
$g_{\mu\nu}=g_{\mu\nu}^{AdS} + 
\sqrt{32\pi G_N } h_{\mu\nu}$, and rescaling the scalar field $\phi \to 
\sqrt{16\pi G_N }\phi$, we obtain canonically normalized kinetic terms $\frac{1}{2} (\nabla \phi)^2$  for the scalar and (\ref{spin2action}) with $M^2=-2$ for $h_{\mu\nu}$.
This means that  the cubic coupling is given by $g_{\phi \phi h_{\mu\nu}} = \sqrt{8\pi G_N}$.
Substituting this value in equation (\ref{Cfromg}) and multiplying by 2 because $\phi_1=\phi_2$,\footnote{Notice that there is an extra Wick contraction in this case.} we indeed confirm the Ward identity (\ref{WardT}) for generic dimension $d$.
 
Let us now specify to the case of planar ${\cal N}=4$ SYM at large 't Hooft coupling $\lambda$ and use its dual description as type IIB superstring theory on AdS$_5\times S^5$.
The string theory action takes the schematic form
\be
S=\frac{1}{2\kappa_{10}^2 } \int d^{10} x \sqrt{g}\,
\left[ 
 \mathcal{L}_{2}+
\ell_s^{2} \mathcal{L}_{4} +\dots
\right],
\ee
where $\mathcal{L}_{k}$ is the part of the Lagrangian density with $k$ spacetime derivatives, $\kappa_{10}$ is the gravitational coupling and $\ell_s$ is the string length.
Expanding around the AdS$_5\times S^5$ background and reducing to AdS$_5$, the effective action becomes
\be
S=\frac{1}{2\kappa^2 } \int d^5 x \sqrt{g}\,
\left[ \ell_s^{-2} \tilde{\mathcal{L}}_{0}+
\tilde{\mathcal{L}}_{2}+
\ell_s^{2} \tilde{\mathcal{L}}_{4} +\dots
\right],
\ee
where $\tilde{\mathcal{L}}_{k}$ contains $k$ or less spacetime derivatives (notice that some derivatives in the ten-dimensional action can act on background fields and produce factors of $1/R$ which is 1 in our units).
The gravitational coupling $\kappa$ satisfies
$\kappa^2 =8 \pi G_N^{(5)}$ and can be removed by rescaling the fields so that they have canonically normalized kinetic terms in the action. This gives 
the following scaling for the  cubic coupling of the type (\ref{eq:bulkinteraction}),
\be
 g_{\phi_1\phi_2h} \sim  \kappa \left[ \ell_s^{J-2} +O\!\left(\ell_s^{J}\right) \right].
 \ee
Converting to gauge theory parameters, the planar OPE coefficient for operators with unit two point function will then be given by 
\be
C_{\phi_1\phi_2h}  \sim
\frac{1}{N}\left(\frac{1}{\lambda}\right)^{\frac{J-2}{4}}\,
 \frac{b(\Delta_1,\Delta_2,\Delta,J)}{\sqrt{\mathcal{C}_{\Delta_1}\mathcal{C}_{\Delta_2}\mathcal{C}_{\Delta,J}}}
\left[1+O\!\left(\frac{1}{\sqrt{\lambda}}\right)\right].
\ee
Note that the explicit dependence on the 't Hooft coupling comes  from the tree level string theory coupling of the
dual fields. There is additionally an implicit dependence on the  't Hooft coupling through the dimension of the operators that
are not protected, in general.
For example, in the case of two protected operators (fixed $\Delta_1$ and $\Delta_2$) and a non-protected operator with $\D(\lambda) \sim \lambda^{1/4}$ and fixed $J$ we find
\be
C_{\phi_1\phi_2h}  \sim
\frac{1}{N} \frac{\lambda^{\frac{\D_1+\D_2-1}{4}} 2^{-\D(\l)}}{
\sin \!\left(\frac{\pi}{2}\big(\D_1+\D_2+J-\D(\lambda)\big) \right)}\,.
\ee

\section{Four-point function}
\label{CPWdecomposition}

As a further application if the embedding formalism we consider four-point functions of scalar primary operators 
computed from Witten diagrams with a spin $J$ field exchange. 
In general, 
a  four-point  function of scalar primary operators in a conformal field theory can be decomposed in partial waves as follows
\cite{Sofia}
\begin{align}
\left\langle \mathcal{O}_{\phi_1}\dots\mathcal{O}_{\phi_4} \right\rangle =\frac{1}{\left(P_{12} \right)^{\frac{\Delta_{1}+\Delta_{2}}{2}}\left(P_{34} \right)^{\frac{\Delta_{3}+\Delta_{4}}{2}}}\left(\frac{P_{24} }{P_{14}}\right)^{\frac{\Delta_{12}}{2}}\left(\frac{P_{14}}{P_{13} }\right)^{\frac{\Delta_{34}}{2}}\sum_{l=0}^{\infty}\int_{-\infty}^{\infty}d\nu \,b_{l}(\nu)\,F_{\nu,l}(u,v)\,,
\label{eq:fourptfunction}
\end{align} 
where the conformal partial wave $F_{\nu,l}(u,v)$ is a function of the cross ratios 
\begin{equation}
 u=\frac{P_{12} P_{34} }{P_{13} P_{24} }\,,   \ \ \ \ \ \ \ \ \ \ \  \ \ \ \ \  \ v=\frac{P_{14}  P_{23} }{P_{13} P_{24} }\,.
 \label{crossratios}
\end{equation}
This function can be defined as the integral of the product of two three-point functions 
\begin{equation}
\!\!
F_{\nu,l}(u,v)=\frac{1}{\beta}
\!
\int\! dP_5 
 \left\langle\mathcal{O}_{\phi_1}(P_1)\mathcal{O}_{\phi_2}(P_2) \mathcal{O}_{h+i\nu,l} (P_5,D_Z)\right\rangle 
 \!
  \left\langle \mathcal{O}_{h-i\nu,l}(P_5,Z) \mathcal{O}_{\phi_3} (P_3)\mathcal{O}_{\phi_4} (P_4)\right\rangle ,\label{eq:CPartialWave} 
\end{equation}
where $\beta$ is a normalization constant given in equation (\ref{Beta}) of appendix \ref{Ap:SplitRepCPW} and  $D_Z$ is the differential operator that implements index contraction defined in (\ref{TodorovOper}). In this expression, the three-point functions are given by
\begin{align}
\left\langle\mathcal{O}_{\phi_1}(P_1)\mathcal{O}_{\phi_2}(P_2)\mathcal{O}_{h+i\nu,l}(P_5,Z) \right\rangle=
\frac{\big((Z\cdot P_1)P_{25}-(Z\cdot P_2)P_{15}\big)^l}
{P_{15}^{\frac{\Delta_1+h+i\nu-\Delta_2+l}{2}}P_{25}^{\frac{\Delta_2+h+i\nu-\Delta_1+l}{2}}P_{12}^{\frac{\Delta_1+\Delta_2+h-i\nu+l}{2}}}\,.
\end{align} 
We remark that the conformal partial wave $F_{\nu,l}(u,v)$ can also be written as a conformal block of dimension 
$h+i\nu$ and spin $l$ plus the conformal block of its shadow operator, which has 
dimension $h-i\nu$ and the same spin.
\footnote{The conformal partial wave $F_{\nu,J}(u,v)$ can be expressed in terms of conformal blocks $G_{\D,J}(u,v)$ as
\begin{align}
F_{\nu,J}(u,v)= \k_{\nu,J}G_{h+i\nu,J}(u,v)+\k_{-\nu,J}G_{h-i\nu,J}(u,v)\,,
\end{align}
where $\k_{\nu,J}$ is a normalization constant defined in   \cite{Costa:2012cb}.}

\begin{figure}
\begin{centering}
\includegraphics[scale=.55]{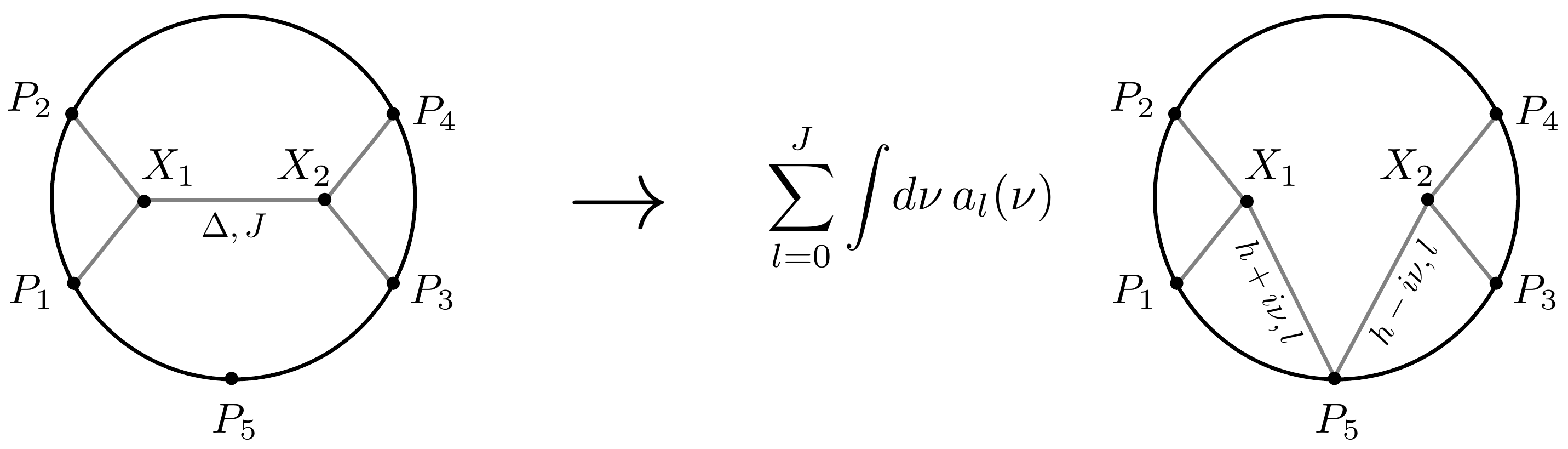}
\par\end{centering}
\caption{\label{fig:Butterfly} Witten diagram describing a spin $J$ exchange between scalar primaries of arbitrary dimension. 
Using the split representation of the bulk-to-bulk propagator this
diagram can be converted into the product of two three-point functions integrated over the common boundary point $P_5$. }
\end{figure}

As an application of the technology developed in the previous sections, we will compute the conformal partial wave decomposition of the four-point function associated to the Witten diagram 
of figure \ref{fig:Butterfly}.
In this diagram, the external operators are 
 scalar fields $\phi_i$, with dimension $\Delta_i$, which exchange a field of dimension $\D$ and spin $J$.
The spin $J$ field couples to the external scalars 
  through the cubic coupling (\ref{eq:bulkinteraction}).
Next we use the split representation of the spin $J$ bulk-to-bulk propagator given by  (\ref{eq:OmegaIntegralBoundary})
and  (\ref{eq:SlipRepStart}), 
\begin{align}
&\frac{g_{\phi_1\phi_2 h}g_{\phi_3\phi_4 h}}{\left[J! \left(\frac{d-1}{2}\right)_J\right]^2}\int dX_1 dX_2\,\Pi_{\Delta_1}(P_1,X_1)
\left[(K_1\cdot\nabla_1)^J
\Pi_{\Delta_2}(P_2,X_1)\right]
\Pi_{\Delta_3}(P_3,X_2)
\label{eq:Split4pt}\\
&
\left[(K_2\cdot \nabla_2)^J 
\Pi_{\Delta_4}(P_4,X_2)\right]
\sum_{l=0}^{J}\int d\nu \, a_l(\nu)
(W_1\cdot\nabla_1)^{J-l}(W_2\cdot\nabla_2)^{J-l}\Omega_{\nu,l}(X_1,X_2;W_1,W_2)
\,,\nonumber
\end{align}
where $g_{\phi_1\phi_2 h}$ and $g_{\phi_3\phi_4 h}$ are the cubic couplings between the external scalars and the spin $J$ field.
The corresponding diagram is also represented in figure \ref{fig:Butterfly}.
The integration over the bulk points  $X_1$ and $X_2$ produces a product of two three-point functions integrated over the point $P_5$ exactly as in the definition of  the conformal partial wave $F_{\nu,J}$. 
Therefore, we conclude that the  coefficients $b_l(\nu)$ of the partial wave expansion  (\ref{eq:fourptfunction}) are proportional to the functions $a_l(\nu)$ of the split representation
(\ref{eq:SlipRepStart2}) of the propagator.
More precisely, we can write 
\begin{align}
b_l(\nu)=g_{\phi_1\phi_2 h}g_{\phi_1\phi_2 h}\,\alpha_l(\nu) \,a_l(\nu)\,,
\label{eq:SplitPartialWaveRelation}
\end{align}
where   $\alpha_l(\nu)$ is independent of $\Delta$ and it is  given in equation (\ref{eq:ExplicitEquationAlpha}) of appendix  \ref{Ap:SplitRepCPW}.

The  structure of the coefficients $b_l(\nu)$ was studied in detail in appendix A.5 of \cite{Costa:2012cb}. In particular, the residues of the spurious poles of $b_l(\nu)$ are related through
\begin{align}
b_l(\nu)\approx&-\frac{\mathcal{Z}_{l+q,q} \, b_{l+q}\big( i(h-1+l)\big)}{\nu^2+(h+l+q-1)^2}\,, \ \ \ \  \ \ \ \ q=1,2,\dots \,,
\label{eq:SpuriousRelationbJ}
\end{align}
where 
\be
\mathcal{Z}_{J,q}= 
   \frac{J!}{ (J-q)! q!}
   \frac{2(-2)^q    \left(\frac{\Delta
   _1+\Delta _2+1-2 h-q}{2}\right)_q 
   \left(\frac{\Delta
   _3+\Delta _4+1-2 h-q}{2}\right)_q 
   \left(\frac{\Delta
   _{12}+1-q}{2}\right)_q 
   \left(\frac{\Delta
   _{34}+1-q}{2}\right)_q}{
   \Gamma (q)
   (h+J-q)_{q-1}}\,.
   \ee
In equation (\ref{eq:SpuriousRelationbJ}) we used the symbol $\approx$ to mean that the two sides of the expression have the same  residue at the pole in $\nu^2$ shown explicitly.
Given  (\ref{eq:SplitPartialWaveRelation}) and (\ref{eq:SpuriousRelationbJ}) it is  possible to derive a similar constraint on the coefficients $a_l(\nu)$, 
\begin{align}
 a_{l}(\nu)&\approx -\frac{\mathcal{Z}_{l+q,q}\,\alpha_{l+q}\big( i(h+l-1)\big)}{\alpha_{l}\big(i(h+q+l-1)\big)}\frac{a_{l+q}\big( i(h-1+l)\big)}{\nu^2+\left(h+l+q-1\right)^2}=
 \nonumber\\
&=\frac{(l+q)!}{l!q!}
\frac{(-1)^{q+1}  }{2^{q-1} (q-1)!  \, (h+l)_{q-1}}\frac{a_{l+q}\big(i(h-1+l )\big)}{\nu^2+\left(h+l+q-1\right)^2}\,,
\label{eq:spuriousal}
\end{align} 
where the last expression was guessed by generalizing the results of explicit calculations for $l=0,\dots,20$ and $q=0,\dots,20$.
It is remarkable that all the dependence of the functions $\mathcal{Z}_{J,q}$, $\a_l$ and $\a_{l+q}$ on the external dimensions $\D_i$  cancelled. This had to happen because $a_l(\nu)$ are the  expansion coefficients of the bulk propagator in the split representation.
This is a very non-trivial consistency check of our results.

In fact, using  (\ref{eq:spuriousal}) we can obtain  the full split representation of a dimension $\D$ and spin $J$ bulk-to-bulk propagator.
The starting point is $a_l(\nu)=0$ for $l>J$ and
\be
a_J(\nu)=\frac{1}{\nu^2+(\D-h)^2}\,,
\ee
as derived in section \ref{SplitRep}. Then, for $l<J$ we take the minimal choice that is compatible with (\ref{eq:spuriousal}),
\begin{align}
 a_{l}(\nu)=\sum_{q=1}^{J-l}
 \frac{(l+q)!}{l!q!}
\frac{(-1)^{q+1}  }{2^{q-1} (q-1)!  \, (h+l)_{q-1}}\frac{a_{l+q}\big(i(h-1+l )\big)}{\nu^2+\left(h+l+q-1\right)^2} \ .
\end{align} 
One can easily check that this reproduces the results  (\ref{eq:splita2}) and (\ref{eq:splita3})
for $l=J-1$ and $l=J-2$.  Moreover, 
one can check that this is consistent with  the explicit expression 
 (\ref{eq:alGenericForm}) for the coefficients $a_l(\nu)$ found in section \ref{subsec:split}.

\subsection{Example: AdS graviton exchange}
One application of the split representation derived above is the computation of the AdS graviton exchange diagram. More precisely, we compute the contribution to the four-point function  of scalar primary operators, from the diagram in 
figure \ref{fig:Butterfly}(a), where the exchanged bulk field is the graviton. This reads 
\footnote{ We include the denominator $\mathcal{C}_{\D_1} \mathcal{C}_{\D_3}$ in order to obtain the four-point function of operators normalized to have unit two-point function (see footnote \ref{2ptnormalization}).}
\begin{align}
&\left\langle \mathcal{O}_{\phi_1}(P_1) \mathcal{O}_{\phi_1}(P_2)\mathcal{O}_{\phi_3}(P_3) \phi_3(P_4) \right\rangle
=
\\
&=\frac{8\pi G_N}{\mathcal{C}_{\D_1} \mathcal{C}_{\D_3}}
\int_{AdS}
dX_1 dX_2 \,T_{AB}^{(12)}(X_1) \,
\Pi_{d,2}^{AB,CD}(X_1,X_2)\,
T_{CD}^{(34)}(X_2) \,,
\nonumber
\end{align}
where
\begin{align}
&T_{AB}^{(12)}(X) = \nabla_A\Pi_{\D_1}(X,P_1)\nabla_B\Pi_{\D_1}(X,P_2)+\nabla_B\Pi_{\D_1}(X,P_1)\nabla_A\Pi_{\D_1}(X,P_2)\\
&-\left(\eta_{AB}+X_{A}X_{B}\right)\Big[\nabla^C\Pi_{\D_1}(X,P_1)\nabla_C\Pi_{\D_1}(X,P_2) +\D_1(\D_1-d)\Pi_{\D_1}(X,P_1)\Pi_{\D_1}(X,P_2)\Big] \,,
\nonumber
\end{align}
and similarly for $T_{CD}^{(34)}(X)$.
This Witten diagram was first computed in \cite{D'Hoker} for some specific values of $\D_1$, $\D_3$ and $d$. Here, we will use the split representation (\ref{SplitGraviton}) of the graviton propagator to obtain directly the conformal partial wave expansion of the Witten diagram for all values of $\D_1$, $\D_3$ and spacetime dimension $d=2h$. This calculation is very similar to the one discussed above for  (\ref{eq:Split4pt}).
Since the the sources $T_{AB}^{(12)}(X_1)$ and $T_{CD}^{(34)}(X_2)$ are conserved we can drop the total derivative terms in (\ref{SplitGraviton}), as expected.
To determine the contribution from the remaining two terms, we use  the 
 representation (\ref{eq:OmegaIntegralBoundary}) of the harmonic functions $\Omega_{\nu,2}$ and $\Omega_{\nu,0}$.
 After integrating over the bulk points $X_1$ and $X_2$, we are left with the integral over the boundary point $P$ in (\ref{eq:OmegaIntegralBoundary}), of the product of two three-point functions like in the definition (\ref{eq:CPartialWave}) of the conformal partial waves. Then the  partial amplitudes,
 as defined in (\ref{eq:fourptfunction}), are given by
 \begin{align}
b_2(\nu)&=\frac{8\pi G_N}{2\pi ^{ h}\G(\D_1)\G(\D_3) \G(\D_1+1-h) \G (\D_3+1-h)}\frac{1}{\nu^2+h^2}\,,
\label{b2ofnu}\\
b_0(\nu)&=-8\pi G_N
\frac{ 
\big[ 4 \D_1(2h-\D_1)+(2
   h-1) (h^2+\nu ^2)\big] 
   \big[ 4 \D_3(2h-\D_3)+(2
   h-1) (h^2+\nu ^2)\big]}{64\pi ^{h}h(2h-1) \Gamma (\D_1+1-h) \Gamma (\D_3+1-h)\G(\D_1)\G(\D_3)\big[\nu^2+(h+1)^2\big]}\,,
   \nonumber
\end{align}     
which determine the conformal partial wave expansion of the graviton exchange diagram.
Notice that this result is consistent with the relation (\ref
{eq:SpuriousRelationbJ}) for the \emph{spurious} poles 
of the partial amplitudes. Moreover,  the pole of $b_2(\nu)$ is fixed by the conformal Ward identity. 
To see that, we first use the relation between the residues of the conformal partial amplitudes $b_l(\nu)$ and OPE coefficients  \cite{Sofia, Mack, Costa:2012cb}. In the particular case of the stress-energy tensor this gives
\begin{align}
\lim_{\nu\to ih} \left(\nu^2+h^2\right) b_2(\nu)=
\frac{h(2h-1)\G(2h+2) \,
C_{\phi_1\phi_1T_{\mu\nu}}C_{\phi_3\phi_3T_{\mu\nu}}
 }{2\G^3(h+1 )\G ( \D_1-h+1 )\G ( \D_3-h+1 )\G ( \D_1 +1 )\G ( \D_3+1 )}\,.
\label{eq:Relationb2OPECoeff}
\end{align}
Secondly, the OPE coefficients are determined by the Ward identity as in (\ref{WardT}). This 
reproduces the relation
(\ref{CTfromGN}) between the bulk gravitational coupling $G_N$ and the CFT central charge $C_T$.


\subsubsection{Mellin amplitude}

Recalling that  given a conformal  four-point function, its Mellin amplitude $M(s,t)$ is defined by
\begin{align}
&\left\langle \mathcal{O}_{\phi_1}(P_1) \mathcal{O}_{\phi_1}(P_2)\mathcal{O}_{\phi_3}(P_3) \mathcal{O}_{\phi_3}(P_4)  \right\rangle = 
\frac{1}{(P_{12})^{\D_1}(P_{34})^{\D_3}}\times\label{eq:StartingMellinAmplitudeGraviton}\\
&\ \ \ \ \ \times\int \frac{dsdt}{(4\pi i)^2}\, 
 M(s,t)\,u^{\frac{t}{2}} v^{-\frac{s+t}{2}}  
 \G\!\left(\D_1-\frac{ t}{2}\right)
  \G\!\left(\D_3-\frac{ t}{2}\right)
\G^2\bigg(\frac{-s}{2}\bigg)\G^2\bigg(\frac{s+t}{2}\bigg)\,,
\nonumber
\end{align}
where the integration contours run parallel to the imaginary axis and the cross-ratios $u$ and $v$ were defined in (\ref{crossratios}).
As explained in \cite{Mack, Costa:2012cb}, the Mellin amplitude also admits a conformal partial wave expansion,
\be
M(s,t) = \sum_{l=0}^\infty \int d\nu \,b_l(\nu)\, M_{\nu,l}(s,t)\,,
\ee
where the partial waves $M_{\nu,l}(s,t)$ involve 
Mack polynomials and are given in \cite{Costa:2012cb} in our conventions.

In the case of the graviton exchange  diagram, we find  the Mellin amplitude can be expressed as the following integral
\begin{align}
&M(s,t)= \int d\nu 
\bigg[P_{\nu,2}(s,t)\,b_2(\nu)
 +\frac{ (t+2-h)^2+\nu ^2 }{\big[(h-2 \D_1)^2+\nu ^2\big]
   \big[(h-2 \D_3)^2+\nu ^2\big]}\,
   4b_0(\nu)
   \bigg]\times\\
   &\  \times \frac{\Gamma \!\left(\frac{2\D_1+2-h-i\nu}{2}\right) \Gamma\! \left(\frac{2\D_1+2-h+i\nu}{2}\right) \Gamma\! \left(\frac{2\D_3+2-h-i\nu}{2}\right) \Gamma\! \left(\frac{2\D_3+2-h+i\nu}{2}\right)\G\!\left(\frac{h-t-i \nu -2}{2}\right) \G\! \left(\frac{h-t+i \nu-2}{2}\right)}{8 \pi  \G\! \left(\frac{2\D_1-t}{2}\right) \G\!\left(\frac{2\D_3-t}{2}\right)\G(i\nu)\G(-i\nu)}\,,
   \nonumber
\end{align}
where the spin 2 Mack polynomial reads
\begin{align}
P_{\nu,2}(s,t)=&\ 
\frac{(2 h-t-3) (2 h-t-1)}
{8h\big[(h-1)^2+\nu ^2\big]}
\\
&\,-\frac{h^2-8 h s^2-8 h s t-2 h t^2
-4 h t-2
   h+\nu ^2+t^2+4 t+3}{8 h}\,.
\nonumber
\end{align}
The integral over $\nu$ can be done explicitly using the following identities, 
\begin{align}
\int d\nu\, \frac{\prod_{\sigma= \pm1}\prod_{k=1}^{3}\G\!\left(\frac{a_k+\sigma i\nu}{2}\right)}{8\pi\G(i\nu)\G(-i\nu)}= \G\!\left(\frac{a_1+a_2}{2}\right)\G\!\left(\frac{a_1+a_3}{2}\right)\G\!\left(\frac{a_2+a_3}{2}\right),
\end{align}
and
\begin{align}
\frac{1}{\G\big(\frac{a-t}{2}\big)\G\big(\frac{b-t}{2}\big)}\int_{-\infty}^{\infty} \frac{d\nu}{4\pi}\frac{l(\nu)l(-\nu)}{\big(\nu^2+(\D-h)^2\big)} =\sum_{m=0}^{\infty}\frac{R_m}{t-\D-2m}\,,
\end{align}
where
\begin{align}
&l(\nu)=\frac{\G\!\left(\frac{h+i\nu-t}{2}\right)\G\!\left(\frac{a+i\nu-h}{2}\right)\G\!\left(\frac{b+i\nu-h}{2}\right)}{ \G(i\nu)}\,,
\nonumber\\
&R_{m}= \G\!\left(\frac{a+\D-2h}{2}\right) \G\!\left(\frac{b+\D-2h}{2}\right) 
\frac{\big(1+\frac{\D-a}{2}\big)_m \big(1+\frac{\D-b}{2}\big)_m}{m!\G\big(\D-h+1+m\big)}\,.
\end{align}
The final  result for generic conformal weights $\D_i$ of the external scalars  and spacetime dimension is
\begin{align}
M(s,t)&=C_{\phi_1\phi_1T_{\mu\nu}}C_{\phi_3\phi_3T_{\mu\nu}}\sum_{m=0}^{\infty}\frac{\mathcal{Q}_{2,m}(s)}{t-2h+2-2m}
\label{finalresultgravitonMellin} \\
&+\frac{8\pi G_N \G
   (\D_1+\D_3-h)
   \big(h s-\D_1 \D_3-s (\D_1+\D_3)\big)}{4 \pi ^{h}\G(\D_1) \G(\D_3) \G(\D_1+1-h) \G(\D_3+1-h)}\,,
   \nonumber
\end{align}
where 
\be
\mathcal{Q}_{2,m}(s)=
\frac{(1-2 h) h \G (2 h+2)P_{ih,2}(s,2h-2+2m)}{4 m!\G^4 (h+1) (h+1)_m \G(\D_1+1-h-m) \G(\D_3+1-h-m)}\,,
\ee
  is the spin 2 Mack polynomial as defined in \cite{Costa:2012cb}  and the product of OPE coefficients  is given by
\be
C_{\phi_1\phi_1T_{\mu\nu}}C_{\phi_3\phi_3T_{\mu\nu}}=8\pi G_{N}\frac{\D_1 \D_3\G^3 (h+1)}{\pi^h h (2h-1) \G (2 h+2)}\,.
\ee
Notice that this value for the OPE coefficient is consistent with (\ref{b2ofnu}) and (\ref{eq:Relationb2OPECoeff}).
The final result (\ref{finalresultgravitonMellin}) for the Mellin amplitude consists of two pieces. The first piece is a sum of poles with residues entirely determined by the product of OPE coefficients $C_{\phi_1\phi_1T_{\mu\nu}}C_{\phi_3\phi_3T_{\mu\nu}}\,$. This piece follows from the structure of the Mellin amplitudes and was known before (see appendix A.3 of \cite{Costa:2012cb}). The second term in (\ref{finalresultgravitonMellin}) was not known before and required a careful treatment of the contact terms in the graviton propagator.

In the large $s \sim t$ limit, the Mellin amplitude simplifies to
\be
M(s,t)\approx -\frac{8\pi G_N \G
   (\D_1+\D_3-h+1)
   }{4 \pi ^{h}\G(\D_1) \G(\D_3) \G(\D_1+1-h) \G(\D_3+1-h)}\frac{s^2+st}{t}\,.
\ee
Using the flat space limit of AdS \cite{JPMellin}, this corresponds to the following bulk scattering amplitude between massless scalars \footnote{See equation (132) of  \cite{Costa:2012cb} for the flat space limit formula in the present conventions.}
\be
\mathcal{T}(S,T)\approx  -8\pi G_N  \,
    \frac{S^2+ST}{T}\,,
\ee
where $S$ and $T$ are the usual flat space  Mandelstam invariants. This is the correct result for the graviton exchange amplitude between minimally coupled massless scalars \cite{FSgraviton1,FSgraviton2}.
Another check of the result (\ref{finalresultgravitonMellin}) is the property $M(s,t)=M(-s-t,t)$, that follows from the invariance of the Witten diagram under the exchange of points $1 \leftrightarrow 2$ (or  $3 \leftrightarrow 4$).
Finally, as an example, we present the result when $\D_1=\D_3=2h=4$,
\begin{align}
M(s,t)=-8\pi G_N\, \frac{1}{12\pi^2}\left(\frac{(s+4) (s+2)}{(t-6)}+\frac{8\left(s+2\right)^2}{ (t-4)}+\frac{6 s(s+2)+8}{(t-2)}+5 (3 s+8)\right),
\end{align}
which matches the result obtained in  \cite{JPMellin} (after the appropriate change of conventions).


The four-point function associated to an AdS graviton exchange between  scalar primary operators can be expanded in conformal blocks in the crossed channel. More precisely, we can write
\be
\left\langle\mathcal{O}_{\phi_1}(P_1) \mathcal{O}_{\phi_1}(P_2)\mathcal{O}_{\phi_3}(P_3) \mathcal{O}_{\phi_3}(P_4)  \right\rangle =
\sum_{n,l=0}^\infty p(n,l) \,G_{\D(n,l),l}^{(13)(24)}(P_1,\dots,P_4)\,,
\ee
where $G_{\D(n,l),l}^{(13)(24)}$ is the conformal block describing the exchange of an operator with spin $l$ and  dimension  
\be
\D(n,l)=\D_1+\D_3+2n+l+\g(n,l)\,,
\ee
in the $(13)(24)$ channel.
The anomalous dimensions $\g(n,l)$ are small if the  interactions in the dual AdS space are weak.
In fact, we can think of $\g(n,l)$ as the gravitational binding energy of a two particle state with angular momentum $l$
\cite{ourEikonal}. The quantum number $n$ increases the ellipticity  of the corresponding classical orbits and we will set it to zero for simplicity.
Using the techniques described in \cite{Costa:2012cb}, one can compute the anomalous dimensions in terms of the Mellin amplitude. 
To first order in the gravitational coupling we obtain
\begin{align}
\g(0,l)=&-\int_{-i\infty}^{i\infty} \frac{dt}{2\pi i} \, M(0,t)\, \G^2\!\left(\frac{t}{2}\right) \G\!\left(\D_1-\frac{t}{2}\right) 
\G\!\left(\D_1-\frac{t}{2}\right)\times
\nonumber
\\
&\ \ \ \ \ \ \ \ \ \ \ \times \ _3F_2\!\left(-l,l+\D_1+\D_3-1,\frac{t}{2};\D_1,\D_3;1\right),
\end{align}
where  the Mellin amplitude was given in
 (\ref{finalresultgravitonMellin}).
Using the large $l$ asymptotic behaviour,
\be
\ _3F_2\left(-l,l+\D_1+\D_3-1,\frac{t}{2};\D_1,\D_3;1\right) \approx
\frac{\G(\D_1)\G(\D_3)}{
\G\left(\D_1-\frac{t}{2}\right)
\G\left(\D_3-\frac{t}{2}\right)}\frac{1}{l^t}\ ,
\ee
we conclude that the large spin behaviour of the anomalous dimension $\g(0,l)$ is controlled the leading $t$-pole of the Mellin amplitude
 (\ref{finalresultgravitonMellin}).
This gives
\be
\g(0,l) \approx -
 C_{\phi_1\phi_1T_{\mu\nu}}C_{\phi_3\phi_3T_{\mu\nu}}\,
 \frac{
\Gamma (2 h+2)\Gamma ( \Delta_1 ) \Gamma ( \Delta_3) }{2 \Gamma^2 (h+1) \Gamma
   (\Delta_1 -h+1) \Gamma (\Delta_3-h+1)}
   \frac{1}{l^{2h-2}}\,,
  \ee
which agrees with the results of \cite{ Fitzpatrick:2012yx, Komargodski:2012ek, Fitzpatrick:2014vua} (in particular see formula (B.33) of \cite{Komargodski:2012ek}), if we assume that  the stress-energy tensor is the operator with minimal twist.

\section{Concluding remarks}

In this work we developed the embedding formalism to deal with tensor fields in Anti-de Sitter spacetime.
In particular, we encoded symmetric traceless tensors into polynomials of an auxiliary null vector and found differential operators that implement the laplacian and divergence in this language.
With this technology, we were able to obtain the bulk-to-bulk propagator of massive spinning particles 
in AdS.  We also found a split representation for these propagators. 
This is a very useful integral representation because it is based on the product of bulk-to-boundary propagators. For example, it leads directly to the conformal partial wave expansion of four-point Witten diagrams.
Up to spin 2, we gave complete split representations of the bulk-to-bulk propagator, including contact terms. 
By a careful study of the massless limit, we obtained the split representation of the graviton propagator which was the subject of some controversy in the literature \cite{Balitsky:2011tw, Giecold:2012qi}.
For spin greater than 2, we only gave the split representation up to contact terms, i.e. for non-coincident points.
It should also be possible to obtain the complete split representation, for instance studying the non-local equations of motion proposed in \cite{Francia:2007ee}.

We illustrated the use of the embedding formalism and the split representation of the propagators, by computing three and four-point functions involving tensor fields.
In particular, we obtained a closed formula for the conformal partial wave expansion and the Mellin amplitude associated to graviton exchange between to scalars of arbitrary conformal weight in general spacetime dimension.

There are several  natural extensions of our work.
An obvious one is to study   antisymmetric and mixed symmetry tensors. More interesting, would be the generalization to spinorial fields.
Another simple extension is the study of higher spin fields in de Sitter space.
 We leave these ideas for the future, hoping  to  
 have convinced the reader that  embedding methods can be very powerful in  the treatment of higher spin fields in AdS, for example in the computation of Witten diagrams.

\section*{Acknowledgements}

We are grateful to Perimeter Institute for the great hospitality in the summer of 2012 where this 
work was initiated.
M.S.C and V.G also thank IPMU at Tokyo University for the great hospitality during the progress of this work.
J.P. wishes to thank the hospitality of KITP at UCSB where part of this work was developed. The research leading to these results has received funding from the [European Union] Seventh Framework Programme under grant agreements No 269217 and No 317089.
This work was partially funded by the grant CERN/FP/123599/2011 and by the Matsumae International Foundation in Japan.
\emph{Centro de Fisica do Porto} is partially funded by the Foundation for 
Science and Technology of Portugal (FCT). The work of V.G. is supported  
by the FCT fellowship SFRH/BD/68313/2010.

\appendix

\section{Harmonic functions  in flat space}
\label{Ap:flatspaceOmega}
In this appendix  we collect some basic facts about harmonic functions in flat space, hopefully this will make the transition to AdS more transparent. 
The equation for the propagator of a spin $J$ field in flat space is simpler than the AdS one (\ref{eq:Laplacian}),
\ba
\left(\partial^2 -m^2 \right)  \Pi_{m}^{A_1\dots A_JB_1\dots B_J}\!\left(X-\bar{X}\right) =-
 \mathcal{P}_{m^2}^{ A_1\dots A_J\,B_1\dots B_J} \delta\!\left(X-\bar{X}\right),
\ea
where $\mathcal{P}_{m^2}$ is a projector. 
For example,
\ba
\mathcal{P}_{m^2}^{ AB} =  \eta^{ A B} - \frac{\partial^A\partial^B }{m^2} \,.
\ea
More generally, the projector is 
\ba
\mathcal{P}_{m^2}^{ A_1\dots A_JB_1\dots B_J} = \frac{1}{J!}\sum_{ {\rm perm}\, B_i}
\prod_{i=1}^J  \mathcal{P}_{m^2}^{ A_iB_i}  -  (A_i{\rm\ and\ }B_i{\rm\ traces})\,.
\ea
The projector commutes with the Laplacian, therefore
\ba
 \Pi_{m}^{A_1\dots A_JB_1\dots B_J}\!\left(X-\bar{X}\right)= \mathcal{P}_{m^2}^{ A_1\dots A_JB_1\dots B_J} 
  \Pi_{m}\!\left(X-\bar{X}\right).
\ea
The analogue of AdS harmonic function (\ref{integrationboundary}) in flat space is
\begin{align}
 \Omega_{\nu}^{A_1\dots A_JB_1\dots B_J}\!\left(X-\bar{X}\right)&=\frac{i\nu}{2\pi}
\left[   \Pi_{i\nu}^{A_1\dots A_JB_1\dots B_J}\!\left(X-\bar{X}\right) -
 \Pi_{-i\nu}^{A_1\dots A_JB_1\dots B_J}\!\left(X-\bar{X}\right) \right]
  \nonumber\\
&=\mathcal{P}_{-\nu^2}^{ A_1\dots A_JB_1\dots B_J} 
 \Omega_{\nu}\!\left(X-\bar{X}\right).
 \end{align}
The harmonic function can be written using an integral representation
\ba
   \Omega_{  \nu}(X) = \nu \int dK \,e^{iK\cdot X}\delta\!\left(K^2-\nu^2\right),
\ea
that can be explicitly checked from 
 \ba
   \Pi_{\pm i \nu}(X) = \int dK \,\frac{e^{iK\cdot X}}{K^2+(\pm i \nu+\epsilon)^2}
   = \int dK\, \frac{e^{iK\cdot X}}{K^2-\nu^2 \pm i \epsilon }\,.
 \ea

The generalization of harmonic functions $\Omega_\nu$ to spin $J$ is now straightforward,
\be
 \Omega_{\nu}^{A_1\dots A_JB_1\dots B_J}(X)= \nu\! \int \!dK \,e^{iK\cdot X}  \delta\!\left(K^2-\nu^2 \right) 
 \frac{1}{J!}
 \sum_{\pi_{B_i}}
\prod_{i=1}^J  \left(\eta^{ A_iB_i} -\frac{K^{A_i}K^{B_i}}{\nu^2} \right) -{\rm traces}\,,
 \ee
 where the sum is over the permutations of all $B_i$ indices.
 For example, for spin $2$ we have 
 \ba
 \Omega_{\nu}^{A_1 A_2B_1 B_2}(X)&=& \frac{\nu}{2} \int dK \,e^{iK\cdot X} \delta\!\left(K^2-\nu^2\right)
 \\
 &&
 \left[
  \left(\eta^{ A_1B_1} -\frac{K^{A_1}K^{B_1}}{\nu^2} \right)   \left(\eta^{ A_2B_2} -\frac{K^{A_2}K^{B_2}}{\nu^2} \right)\right.
  \nonumber\\
  &&+\left(\eta^{ A_1B_2} -\frac{K^{A_1}K^{B_2}}{\nu^2} \right)   \left(\eta^{ A_2B_1} -\frac{K^{A_2}K^{B_1}}{\nu^2} \right)
 \nonumber\\
 &&\left.-\frac{2}{d}
 \left(\eta^{ A_1A_2} -\frac{K^{A_1}K^{A_2}}{\nu^2} \right)   \left(\eta^{ B_1B_2} -\frac{K^{B_1}K^{B_2}}{\nu^2} \right)\right],
  \nonumber
  \ea
with $d+1$ being the spacetime dimension.

These harmonic functions satisfy orthogonality and completeness relations similar to   (\ref{eq:ConstraintFinal}) and (\ref{eq:CompletenessNice}). For scalars, we have
\ba
\int d\nu \, \Omega_{\nu}\!\left(X-\bar{X}\right) =  \delta\!\left(X-\bar{X}\right),
\label{eq:Completeness1FlatspaceScalar}
\ea
and
\begin{align}
\int dY \,\Omega_{\nu}\!\left(X-Y\right) \Omega_{\bar{\nu}}\!\left(Y-\bar{X}\right) 
= \frac{\d(\nu+\bar{\nu})+\d(\nu+\bar{\nu})}{2}\,\Omega_{\nu}\!\left(X-\bar{X}\right).
\label{eq:Completeness2FlatspaceScalar}
\end{align}
The generalization of (\ref{eq:Completeness2FlatspaceScalar}) to non-zero spin is simply 
\begin{align}
&\int dY\,\Omega_{\nu}^{A_1\dots A_J B_1\dots B_J}(X-Y)\,\Omega_{\bar{\nu}}^{B_1\dots B_JC_1\dots C_J}\!\left(Y-\bar{X}\right) 
\nonumber\\
&= \frac{\d(\nu+\bar{\nu})+\d(\nu+\bar{\nu})}{2} \, \Omega_{\nu}^{A_1\dots A_J\, C_1\dots C_J}\!\left(X-\bar{X}\right).
\label{eq:Completeness2FlatspaceSpinJ}
\end{align}
On the other hand, the generalization of (\ref{eq:Completeness1FlatspaceScalar}) to generic spin is more subtle. For spin 1 and 2 one has
\ba
 \int d\nu  \left[ \Omega^{AB}_{\nu}\!\left(X-\bar{X}\right) +\frac{1}{\nu^2} \,\partial^A\bar{\partial}^B\Omega_{\nu}\!\left(X-\bar{X}\right)\right]= \eta^{AB} \delta\!\left(X-\bar{X}\right),\label{eq:CompletenessRelation1spin1}
  \ea
 and    
\begin{align}
 \int d\nu &  \left[ \Omega^{A_1A_2B_1B_2}_{\nu}\!\left(X-\bar{X}\right) + \frac{1}{2\nu^2}\,\partial^{A_1}\bar{\partial}^{B_1}\Omega^{A_2B_2}_{\nu}\!\left(X-\bar{X}\right) 
 +\frac{1}{2\nu^2} \,\partial^{A_2}\bar{\partial}^{B_2}\Omega^{A_1B_1}_{\nu}\!\left(X-\bar{X}\right)
 \right. 
 \nonumber
 \\&
 +\frac{1}{2\nu^2} \,\partial^{A_1}\bar{\partial}^{B_2}\Omega^{A_2B_1}_{\nu}\!\left(X-\bar{X}\right)
+\frac{1}{2\nu^2} \,\partial^{A_2}\bar{\partial}^{B_1}\Omega^{A_1B_2}_{\nu}\!\left(X-\bar{X}\right)
 \nonumber
 \\&
 \left.
 + \frac{d+1}{d\,\nu^4}\, \partial^{A_1}\partial^{A_2}\bar{\partial}^{B_1}\bar{\partial}^{B_2}\Omega_{\nu}\!\left(X-\bar{X}\right)
 +\frac{1}{d}\,\eta^{ A_1A_2} \eta^{ B_1B_2}\Omega_{\nu}\!\left(X-\bar{X}\right)
 \right]
  \nonumber
  \\&
=  \frac{1}{2}\left(\eta^{ A_1B_1}\eta^{ A_2B_2}+\eta^{ A_1B_2}\eta^{ A_2B_1} \right) \delta\!\left(X-\bar{X}\right),
\end{align}
respectively.
This should be compared with the completeness formula (\ref{eq:CompletenessNice}) for harmonic functions in AdS. In fact, the limit $\nu \gg 1$ of (\ref{eq:CompletenessNice}) matches exactly the flat space formulas above.


\section{Embbeding space operations}
\label{Ap:EmbeddingOperations}

The embedding formalism gives a compact form to write all tensor structures  of the spin $J$ propagators in AdS and  it allows simpler tensor manipulations. The main aim of this section is to compute the trace over all structures present in the propagator, 
\begin{align}
(K_1\cdot K_2)^J(W_{12})^{J-l}\big((W_1\cdot X_2)( W_2\cdot X_1)\big)^{l}\label{eq:TraceStruc}.
\end{align}
To make the derivation more pedagogical, we will present two intermediate tensor operations that are necessary to  compute the trace (\ref{eq:TraceStruc}).

\noindent
{\bf First operation - symmetric traceless contraction:} The first operation is the action of the differential operator $K_1$ on the $W_1$ variables\footnote{In this process we consider $K_2$ as a generic vector. To compute the trace (\ref{eq:TraceStruc}), $K_2$ needs to act on the variables $W_2$ and so we have to remember that the differential operator $K_2$ has to be placed to the left of all vectors $W_2$. }. For that we will use the identity,
\begin{equation}
\frac{(K\cdot P)^J(W\cdot Q)^J}{J!\left(h-\frac{1}{2}\right)_J} =B^{\{ A_1}\dots B^{A_J\}}G_{\{A_1}\dots G_{A_J\}}=\frac{J!\left(B^2G^2\right)^{\frac{J}{2}}}{2^J\left(h-\frac{1}{2}\right)_J}\,C_{J}^{h-\frac{1}{2}}(t)\,,\label{eq:ImportantTensorManipulation}
\end{equation}
where $C_{J}^{h-\frac{1}{2}}(t)$ is the Gegenbauer polynomial and 
\begin{align}
B_{A}=P_{A}+ (X\cdot P) X_{A}\,, \ \ \   G_{A}= Q_{A}+(Q\cdot X) X_{A} \,, 
\  \ \ \ 
  t=\frac{P\cdot Q+(P\cdot X)( Q\cdot X)}{(P\cdot X) \big((Q\cdot X)^2+ Q^2 \big)^{\frac{1}{2}}}\,,
\end{align}
where it was used that $P^2=0$. 
This last expression will also be useful to compute (\ref{eq:Kaction2}) below.

\noindent
{\bf Second operation:} The second operation is the action of the operator $K$ on a specific vector containing the polarization vector $W$,
\be
K_{A_1}\dots K_{A_J}\big(W_{A_1}+\alpha Y_{A_1}(W\cdot Y)\big)\dots \big(W_{A_J}+\alpha Y_{A_J}(W\cdot Y)\big)\,,
\label{eq:SecondTensorOperation}
\ee
where $\alpha$ is any function that does not depend on $W$, and $Y$ is a generic bulk vector satisfying $Y^2=-1$. The result of this operation is
\be
(2h-1)_{J}\left(h+\frac{1}{2}\right)_J\,_2F_1\!\left(1,-J,h+\frac{1}{2},\frac{\alpha (1+X\cdot Y)(1-X\cdot Y)}{2}\right).
\ee
This expression was guessed by performing the operation for $J=1,\,\dots,5$ and should be valid for any  $J$.

\noindent
{\bf Trace over structures of propagator:}
Multiplying (\ref{eq:TraceStruc}) by $\frac{J!x^l}{l!(J-l)!}$ and summing over $l$ gives a generator of all the structures,
\begin{align}
(K_1\cdot K_2)^J\Big(W_1\cdot \big(W_2+x  ( W_2\cdot X_1)X_2\big)\Big)^{J}\,.
\label{eq:TraceStrucGen}
\end{align}
 The original expression can be recovered expanding the final result in $x$. 
Applying  (\ref{eq:ImportantTensorManipulation}) to this expression, we obtain
\be
\sum_{k=0}^{\left|J/2\right|} 
\frac{\theta^k(J!)^2\!\left(h-\frac{1}{2}\right)_{\!J-k} }{2^{2k}k!(J-2k)!}\,
(K_2\cdot X_1)^{2k}\Big(K_2\cdot \big(W_2 + \alpha X_1(W_2\cdot X_1) \big) \Big)^{J-2k} (W_2\cdot X_1)^{2k}
\,,
\label{eq:secondStep}
\ee
where $\left|J/2\right|$ is the integer part of $J/2$,   $\alpha=1+x (X_1 \cdot X_2)$ and $\theta=x^2-\alpha^2 $.  In this calculation we
have used   $K_2\cdot \big(x X_2 (X_1\cdot W_2)\big)=0$. 
The evaluation of the expression outside the fraction in 
(\ref{eq:secondStep}) can be done by multiplying by $2\frac{J!}{(2k)!(J-2k)!}y^{2k}$ and summing over $k$, giving 
\be
\Big(K_2\cdot\big(W_2+(\alpha+y)X_1(W_2\cdot X_1)\big)\Big)^J+\Big(K_2\cdot\big(W_2+(\alpha-y)X_1(W_2\cdot X_1)\big)\Big)^J\,.
\ee
Then, using the second tensor operation (\ref{eq:SecondTensorOperation}) we obtain
\begin{align}
2(2h-1)_{J}\left(h+\frac{1}{2}\right)_J\sum_{m=0}^J\sum_{k=0}^{\left|m/2\right|}\frac{(-J)_m(-1)^m}{2^m\left(h+\frac{1}{2}\right)_m}\frac{m!\big(u(2+u)\big)^m\alpha^{m-2k}y^{2k}}{(m-2k)!(2k)!}\,.
\end{align}
Notice that, as expected, this expression just depends on even powers of $y$. Thus, the expression outside the fraction in 
(\ref{eq:secondStep})  is  recovered by matching powers of $y$, 
\begin{align}
\frac{(2h-1)_{J}\left(h+\frac{1}{2}\right)_J}{J!}\sum_{m=0}^J\frac{(-J)_{m}(-1)^m}{2^m\left(h+\frac{1}{2}\right)_m}\frac{m!\big(u(2+u)\big)^m\alpha^{m-2k}}{(1+J-2k)_{m-J}}\,. 
\end{align}
Joining all pieces in a single expression we conclude that
\begin{align}
&\frac{1}{J!(2h-1)_{J}\left(h+\frac{1}{2}\right)_J}(K_1\cdot K_2)^J \big(W_{12}+x (W_1\cdot X_2)( W_2 \cdot X_1)\big)^J =
\\
&
\!\sum_{k=0}^{\left|J/2\right|}\sum_{m=0}^J \frac{\left(h-\frac{1}{2}\right)_{J-k}}{2^{2k}k!}
\frac{(-J)_{m}m!\big(u(2+u)\big)^m\big(2(1+u)x-1-u(2+u)x^2\big)^k
\big(x(1+x)-1\big)^{m-2k}
}{2^m\left(h+\frac{1}{2}\right)_m(m-2k)!}\,.
\nonumber
\end{align}  
This simplifies dramatically in the limit $u\rightarrow 0$, as only the $x=0$ term survives. The final result is
\be
(K_1\cdot K_2)^J \big(W_{12}+x (W_1\cdot X_2 )(W_2 \cdot X_1)\big)^J=(2h-1)_{J}\left(h+\frac{1}{2}\right)_{\!\!J}\left(h-\frac{1}{2}\right)_{\!\!J}J! +O(u)\,.
\ee

\section{Split representation of the  bulk-to-bulk propagator}\label{Ap:BulkToBoundary}

 
The harmonic functions $\Omega_{\nu,J}$ can be defined by the integral over the boundary  point that connects two bulk-to-boudary propagators, as in equation (\ref{eq:OmegaIntegralBoundary}).
Alternatively, the harmonic functions $\Omega_{\nu,J}$ can be defined by the difference of two bulk-to-bulk propagators with dimensions $h+i\nu$ and  $h-i\nu$, as in equation  (\ref{integrationboundary}).
The goal of this appendix is to show that these are equivalent definitions.

We start from equation (\ref{eq:OmegaIntegralBoundary}),
\begin{align}
\frac{\nu^2}{\pi  J!(h-1)_J}\,
\int_{\partial} dP \,\Pi_{h+i\nu,J}(X_1,P;W_1,D_Z)\,\Pi_{h-i\nu,J}(X_2,P;W_2,Z)\,.
\end{align}
The boundary contraction can be done using an identity similar to (\ref{eq:ImportantTensorManipulation}) \cite{SpinningCC,Costa:2011dw},
with result
\be
\frac{2^JJ!\mathcal{C}_{h+i\nu,J} \mathcal{C}_{h-i\nu,J}\nu^2}{\pi(h-1)_J}\int_{\partial} dP \,\frac{\big( (P\cdot W_1)( P\cdot W_2)\big)^J C_J^{h-1}(t)}{\left(-2P\cdot X_1\right)^{h+i\nu+J}\left(-2P\cdot X_2\right)^{h-i\nu+J}}\,,\label{eq:BulkToBoundarYIntegral}
\ee 
where t is defined as
\be
t=X_1\cdot X_2 +\frac{ (P\cdot X_1 )(P\cdot X_2)}{(P\cdot W_1 )(P\cdot W_2)} \, W_1\cdot W_2
-\frac{P\cdot X_1}{P\cdot W_1} \,W_1\cdot X_2-\frac{P\cdot X_2}{P\cdot W_2}\,W_2\cdot X_1\,.
\ee
It is possible to choose polarizations such that $W_1\cdot X_2$  and $W_2\cdot X_1$ vanish. With this specific choice only the term in $\left(W_{12}\right)^J$ survives. 
Using the definition of the Gegenbauer polynomial and performing a Feynman parametrization, (\ref{eq:BulkToBoundarYIntegral}) becomes
\begin{align}
&\frac{J!\mathcal{C}_{h+i\nu,J}\mathcal{C}_{h-i\nu,J}\nu^2}{\pi(h-1)_J}\sum_{l=0}^{J}\sum_{k=0}^{\left|l/2\right|}\int_{\partial} dP \int_0^\infty \frac{dq}{q} \,\times
\label{interm_step}
\\
&\times
\frac{(-1)^{k+l}2^{l}q^{h+l}\,\G(2h+2l)\big(h-1\big)_{J-k} (W_{12})^{J-l}\big( (W_1\cdot P)( W_2\cdot P)\big)^{l}(2 \,X_1\cdot X_2)^{l-2k}}
{k! (l-2k)! (J-l)! q^{i\nu}\,\Gamma(h+i\nu+l) \,\Gamma(h-i\nu+l)\left(-2P\cdot Y\right)^{2h+2l}}\,,
\nonumber
\end{align}
with $Y=X_1+qX_2$.
The integral over the boundary point $P$ is conformal and can be done using the equality \cite{SimmonsDuffin:2012uy}
\be
\int_{\partial} dP\,\frac{P^{A_1} \dots P^{A_{2l}}}{\left(-2P\cdot Y\right)^{2h+2l}}=\frac{\pi^h\left(2h+2l\right)_{-h}Y^{A_1} \dots Y^{A_{2l}}}{\left(-Y^2\right)^{h+2l}}
-{\rm traces}\,.
\ee
As the integration variables are contracted with $W_1$ and $W_2$, and since $W_i\cdot X_j=0$, we have
\be
\int_{\partial} dP\,\frac{(W_1\cdot P)^l(W_2\cdot P)^l}{(-2P\cdot Y)^{2h+2l}} =
\frac{\pi^h\left(2h+2l\right)_{-h}l!}{\left(-2\right)^l\left(h+l\right)_{l}\left(-Y^2\right)^{h+l}}\left(W_{12}\right)^{l}\,.
\ee
The sum over $k$ can be done and gives an hypergeometric function, so (\ref{interm_step}) becomes
\begin{align}
\frac{\mathcal{C}_{h+i\nu,J}\mathcal{C}_{h-i\nu,J}\nu^2}{\pi} &\sum_{l=0}^{J}\int \!
dq \,\frac{\pi^hJ!\,\G(h+l)(W_{12})^J \,\!\ _2F_1\!\left(-l,\frac{3}{2}-h-J,3-2h-2J,-\frac{2}{u}\right)}
{\left(J-l\right)! \,\G(h+i\nu+l)\,\G(h-i\nu+l) \left(-2u\right)^{-l}q^{1-h+i\nu-l}\left(-Y^2\right)^{h+l}}\,.
\end{align}
Finally,
using the equality
\begin{align}
\int_{0}^{\infty}\frac{d\alpha}{\alpha}\frac{\alpha^{-c}}{\left(\frac{\left(1+\alpha\right)^2}{\alpha}+2u\right)^b}=\ &
\frac{\Gamma(b+c)\,\Gamma(-c)}{\Gamma(b)\left(2u\right)^{b+c}}\,\!\ _2F_1\!\left(\frac{1}{2}+c,b+c,1+2c,-\frac{2}{u}\right)+\nonumber\\
&+\frac{\Gamma(b-c)\,\Gamma(c)}{\Gamma(b)\left(2u\right)^{b-c}}\,\!\ _2F_1\!\left(\frac{1}{2}-c,b-c,1-2c,-\frac{2}{u}\right),
\end{align}
the integral over $q$ can also be expressed in terms of hypergeometric functions.
These two hypergeometric functions correspond precisely to the two propagators with $\Delta=h\pm i\nu$ in (\ref{integrationboundary}).
Thus, from the term proportional to 
$(W_{12})^J$ in the propagator (\ref{eq:propagatorJgk}), we obtain 
\begin{align}
&g_0(u)= \frac{2\pi^h J!\mathcal{C}_{\D,J}\mathcal{C}_{2h-\D,J }
\G(1+h-\D)
}{ \G(2h -\D)\left(2u\right)^{\D}
}\times\\
&\times\!\sum_{l=0}^{J}\frac{(-1)^l
\ _2F_1 \!\left(-l,\frac{3}{2}-h-J,3-2h-2J,-\frac{2}{u}\right) \,_2F_1\!\left(\frac{1}{2}+\D-h,\D+l,1 +2\D-2h,-\frac{2}{u}\right) }
{\left(J-l\right)! \left(2h-\D\right)_l }\,.
 \nonumber
\end{align} 
We checked up to $J=5$ that this expression for $g_0$ is reproduced by formula (\ref{eq:equacao entre g e f}) where the functions $f_i^{(i+k)}=h_i^{(k)}$ are determined by (\ref{eq:solution}) 
and (\ref{eq:h0}). This shows that (\ref{integrationboundary}) is equivalent to (\ref{eq:OmegaIntegralBoundary}).

\section{AdS harmonic functions}
\label{Ap:orthoOmega}

\subsection*{Orthogonality}

\begin{figure}
\begin{centering}
\includegraphics[scale=0.55]{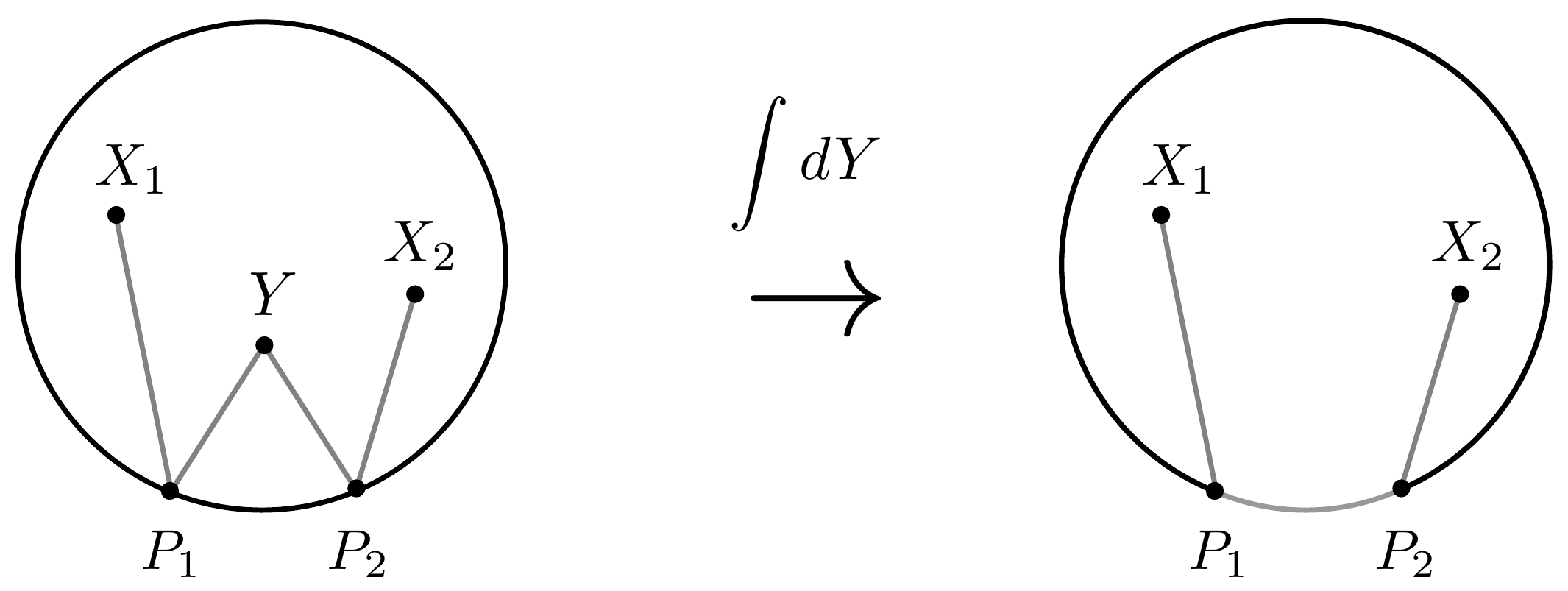}
\par\end{centering}
\caption{\label{fig:Completeness} Product of two  harmonic functions $\Omega_{\nu,J}(X_1,Y;W_1,K)$ and   $\Omega_{\bar{\nu},J}(Y,X_2;W,W_2)$,
where $K$ acts on the polarisation vector $W$ of $Y$,
represented as the integral over bulk-to-boudary propagators. After integration over the bulk point $Y$, we are left with two bulk-to-boudary propagators and one two-point function on the boundary. 
}
\end{figure}

The AdS harmonic functions satisfy the orthogonality relation (\ref{eq:ConstraintFinal}) that says that the integral of two $\Omega$ functions over a commom bulk point gives again an $\Omega$ function. The argument that led to the orthogonality relation could not fix the overall constant multiplying the right hand side of (\ref{eq:ConstraintFinal}). 
The goal of this appendix is to fix this constant by evaluating this bulk integral using  the representation (\ref{eq:OmegaIntegralBoundary}) of the AdS harmonic functions. This computation is represented in figure \ref{fig:Completeness}. We need to do the following integral
\begin{align}
\frac{1}{J!\left(h-\frac{1}{2}\right)_J}
\left(\frac{\nu \bar{\nu} }{\pi J! (h-1)_J}\right)^2
&\int dP_1dP_2 dY \, \Pi_{h+i\nu,J}\!\left(X_1,P_1;W_1,D_{Z_1}\right) \Pi_{h-i\nu,J}\!\left(Y,P_1;K,Z_1\right) \times \nonumber\\
&\ \times\Pi_{h-i\bar{\nu},J}\!\left(Y,P_2;W,D_{Z_2}\right) \Pi_{h+i\bar{\nu},J}\!\left(X_2,P_2;W_2,Z_2\right),
\label{eq:StartingPoint}
\end{align}
where the operator $K$ and $D_{Z}$ were included to perform the index contraction between the propagators.  
We start by performing the integral over the bulk point $Y$,
\begin{align}
\frac{1}{J!\left(h-\frac{1}{2}\right)_J}
&\int  dY \, \Pi_{h-i\nu,J}\!\left(Y,P_1;K,Z_1\right)  
 \Pi_{h-i\bar{\nu},J}\!\left(Y,P_2;W, Z_2 \right)  .
\end{align}
After the contraction over the polarizations, this gives \footnote{Notice that generally contraction of symmetric and traceless structures gives a Gegenbauer polynomial. In this particular case the expression is simplified because the denominator of $t$ in (\ref{eq:ImportantTensorManipulation}) vanishes.}
\begin{align}
&
(-4)^{J}\mathcal{C}_{h-i\nu,J}\mathcal{C}_{h-i\bar{\nu},J}
\int dY \, 
I_{\nu,\bar{\nu},J}\,,
\label{eq:Contraction2BtB}
\end{align}
where  
\begin{align}
&I_{\nu,\bar{\nu},J} =
\frac{1}{(-2P_1\cdot Y)^{h-i\nu+J}(-2P_2\cdot Y)^{h-i\bar{\nu}+J}}
\Big(  (P_1\cdot Z_2)(P_2\cdot Y)(Y \cdot Z_1)
 \\
&+(P_2\cdot Z_1)(P_1\cdot Y )(Y\cdot Z_2)- 
(P_1\cdot P_2)(Y\cdot Z_1)(Y\cdot Z_2)-
(P_1\cdot Y)( P_2\cdot Y )(Z_1\cdot Z_2) \Big)^J\,.
\nonumber
\end{align}
and  the constant $\mathcal{C}_{\D,J}$ is defined in (\ref{eq:normalizationboundary}).
Next we define the differential operator 
\begin{align}
\mathcal{D} = \ &D_{21}D_{12}+D_{12}D_{21}-H_{12}\Big( (P_1\cdot\partial_{P_1})(P_2\cdot\partial_{P_2})+ \frac{1}{2}(P_1\cdot\partial_{P_1}+P_2\cdot\partial_{P_2})\nonumber\\
&-2 (Z_1\cdot\partial_{Z_1})(P_1\cdot\partial_{P_1}+P_2\cdot\partial_{P_2})-2Z_1\cdot\partial_{Z_1} +4 (Z_1\cdot\partial_{Z_1})(Z_1\cdot\partial_{Z_1})\Big)\,,
\end{align}
where
\begin{align}
H_{12}&=2\big( (P_1\cdot Z_2 )(P_2\cdot Z_1)-(P_1\cdot P_2)( Z_1\cdot Z_2)\big)\,,
\label{H_structure}\\
D_{ij}&=  (Z_j\cdot P_i) \left( Z_j\cdot \frac{\partial}{\partial Z_j}-P_j\cdot\frac{\partial}{\partial P_j} \right) + (P_j\cdot P_i)\left(  Z_j\cdot \frac{\partial}{\partial P_j}\right).
\end{align}
Then we have
\be
\frac{1}{4(h-i\nu+J)(h-i\bar{\nu}+J)} \,\mathcal{D} \left( (P_{12})^J I_{\nu,\bar{\nu},J}\right) 
=(P_{12})^{J+1}I_{\nu,\bar{\nu},J+1}\,.
\ee
Thus, the differential operator $\mathcal{D}$ allows us to write
\be
I_{\nu,\bar{\nu},J}=\frac{1}{4^{J}(h-i\nu)_J(h-i\bar{\nu})_J(P_{12})^J}\,\mathcal{D}^J\frac{1}{(-2P_1\cdot Y)^{h-i\nu}(-2P_2\cdot Y)^{h-i\bar{\nu}}}\,.
\ee
So the integral (\ref{eq:Contraction2BtB}) can be rewritten as
\begin{align}
\frac{\mathcal{C}_{h-i\nu,J}\mathcal{C}_{h-i\bar{\nu},J}}{(h-i\nu)_J(h-i\bar{\nu})_J(P_{12})^J} \,\mathcal{D}^{J}\int dY \,\frac{1}{(-2P_1\cdot Y)^{h-i\nu}(-2P_2\cdot Y)^{h-i\bar{\nu}}} \,.
\label{eq:SimplifyingIntegralBulkIndices}
\end{align} 

Now the integral over $Y$ only involves scalar bulk-to-boudary propagators. So, let us analyze
\begin{align}
&
\int \frac{dY}{(-2P_1\cdot Y)^{h-i\nu}(-2P_2\cdot Y)^{h-i\bar{\nu}}} =
\nonumber
\\&
=\int  d^dx\int_0^\infty \frac{dz}{z^{d+1}}\,\frac{ z^{2h-i\nu-i\bar{\nu}+\epsilon}}{\big[z^2+(x-x_1)^2\big]^{h-i\nu}\big[z^2+(x-x_2)^2\big]^{h-i\bar{\nu}}}\,,\label{eq:ScalarIntegralBulkCommomPoint}
\end{align}
where the bulk-to-boudary propagator was written in Poincar\'e coordinates and we have introduced a regulator $z^{\epsilon}$, planning to take the limit $\epsilon \to 0$ at the end. Using the Fourier representation of the propagator, we obtain
\begin{align}
 \int_{0}^{\infty} \frac{dz}{z}\int \frac{d^dk}{(2\pi)^d}  \,z^{-i\nu-i\bar{\nu}+\epsilon} \,e^{ik\cdot(x_1-x_2)}\,\hat{\Pi}_{i\nu}(|k|z) \, \hat{\Pi}_{i\bar{\nu}}(|k|z)\,,\label{eq:Bulktoboundarybulkintegrated}
\end{align}
where the function $\hat{\Pi}_{i\nu}(|k|z)$ is just the Fourier transform
\begin{align}
&\int d^dx \frac{e^{ik\cdot x}}{(z^2+x^2)^{h-i\nu}}=
\\
&=
\int_{0}^{\infty}\frac{ds}{s} \,s^{h-i\nu}\,\frac{e^{-sz^2}e^{-\frac{k^2}{4s}}}{\G(h-i\nu)}
\int d^dx \,e^{-s\left(x-i\frac{k}{2s}\right)^2}=\frac{\pi^h}{\G(h-i\nu)}\int_{0}^{\infty}\frac{ds}{s}\,s^{-i\nu}\,e^{-sz^2-\frac{k^2}{4s}}\,.
\nonumber 
\end{align}
This turns  (\ref{eq:Bulktoboundarybulkintegrated}) into
\begin{align}
\frac{\pi^{2h}}{\G(h-i\nu)\G(h-i\bar{\nu})}
  \int \frac{d^dk}{(2\pi)^d}\,e^{ik\cdot(x_1-x_2)}
\bigg(\frac{4}{k^2}\bigg)^{\frac{i\nu+i\bar{\nu}}{2}}
\int \frac{dz}{z}\frac{dsdt}{st}\,z^{\epsilon}s^{-i\nu}t^{-i\bar{\nu}}e^{-z\big(s+t+\frac{1}{s}+\frac{1}{t}\big)}\,,
\label{eq:Next178}
\end{align}
where  the dependence on $k$ was brought into an explicit form by doing a rescaling $s\rightarrow \frac{sk^2}{4z}$, $t\rightarrow  \frac{tk^2}{4z}$ and $z\rightarrow \frac{2z}{k}$. The  second integral in (\ref{eq:Next178}) is
\begin{align}
&\int \frac{dz}{z}\frac{dsdt}{st} \, z^{\epsilon}s^{-i\nu}t^{-i\bar{\nu}}e^{-z\big(s+t+\frac{1}{s}+\frac{1}{t}\big)} 
=2\G(\epsilon)\int_{-\infty}^{\infty}  dU dV\, \frac{e^{-iU(\nu+\bar{\nu})-iV(\nu-\bar{\nu})}}{\big(4\cosh(U) \cosh(V)\big)^{\epsilon}}=
\nonumber\\
&=\frac{\prod_{n,m=\pm1}\G\big(\textstyle{\frac{\epsilon+i m  \nu+in\bar{\nu} }{2}}\big)}{2\G(\epsilon)}\underset{\epsilon \rightarrow 0}{=}2 \pi\G(i\nu)\G(-i\nu) 
\big[\,\d\!\left(\nu-\bar{\nu}\right)+\d\!\left(\nu+\bar{\nu}\right)\big]\,,
\end{align}
where $U$ and $V$ are related to $t$ and $s$ by  $s=e^{U+V}$ and $t=e^{U-V}$. We conclude
that the integral (\ref{eq:ScalarIntegralBulkCommomPoint}) over AdS of two scalar bulk-to-boudary propagators  is
\begin{align}
&\frac{2\pi^{2h+1}\G(i\nu)\G(-i\nu)}{\Gamma(h-i\nu)\G(h-i\bar{\nu})}
\int \frac{d^dk}{(2\pi)^d} \,e^{ik\cdot(x_1-x_2)}\bigg(\frac{k^2}{4}\bigg)^{-\frac{i\nu+i\bar{\nu}}{2}}\big[\d(\nu-\bar{\nu})+\d(\nu+\bar{\nu})\big]=
\nonumber\\
&=\frac{2\pi^{h+1}\G(-i\nu)}{\G(h-i\nu)(x_{12}^2)^{h-i\nu}}\,\d\!\left(\nu-\bar{\nu}\right)
+\frac{2\pi^{2h+1}\G(i\nu)\G(-i\nu)}{\G(h+i\nu)\G(h-i\nu)}\,\d^d(x_1-x_2)\d\!\left(\nu+\bar{\nu}\right).
\nonumber
\end{align}
Now we just have to act $J$ times with the differential operator $\mathcal{D}$ according to (\ref{eq:SimplifyingIntegralBulkIndices}). Notice that
\begin{align}
-\frac{1}{(J+\D-1)(J+\D)}\,\mathcal{D}\,
\frac{(H_{12})^J}{(P_1 \cdot P_2)^{\D}}=\frac{(H_{12})^{J+1}}{(P_1 \cdot P_2)^{\D}},
\end{align}
where the structure $H_{12}$ is defined in (\ref{H_structure}),
so $\mathcal{D}$ will generate the structure of the spin $J$ boundary two-point function,
\begin{align}
\frac{(-1)^J}{(\D-1)_J(\D)_J}\,
\mathcal{D}^J\,\frac{1}{(-2P_1\cdot P_2)^{\D}}=\frac{(H_{12})^{J}}{(-2P_1\cdot P_2)^{\D}}\,.
\end{align}
Finally, we conclude that 
\begin{align}
&\frac{1}{J!\left(h-\frac{1}{2}\right)_J}
\int  dY \, \Pi_{h-i\nu,J}(Y,P_1;K,Z_1)\,  
 \Pi_{h-i\bar{\nu},J}(Y,P_2;W, Z_2 )  =
 \nonumber\\
=\ &\mathcal{C}_{h-i\nu,J}\mathcal{C}_{h-i\bar{\nu},J}\bigg[\d\!\left(\nu-\bar{\nu}\right)\frac{2\pi^{h+1}(h-i \nu -1) \Gamma (-i \nu )
}{ (h+J-i\nu -1) \Gamma (h-i \nu )}\,\frac{(H_{12})^J}{(P_{12})^{h-i\nu+J}}
\nonumber\\
&+\d\!\left(\nu+\bar{\nu}\right)
\frac{J!2\pi^{2h+1}\G(i\nu)\G(-i\nu)}{(h-i\nu)_J(h-i\bar{\nu})_J\G(h+i\nu)\G(h-i\nu)}\,\frac{\mathcal{D}^J\d^{d}(P_1,P_2)}{(P_{12})^J} \bigg]
\,.
\label{eq:CompleteBulkIntegralSpinJ}
\end{align}

There remains one integral to do on the boundary, say on the point $P_2$. Let us start by performing the integral of the term proportional to $\d(\nu-\bar{\nu})$ in (\ref{eq:CompleteBulkIntegralSpinJ}), 
\begin{align}
&\int dP_2\, \frac{\Pi_{h+i\nu,J}(X_2,P_2;W_2,D_{Z_2})(H_{12})^J }{J! (h-1)_J(P_{12})^{h-i\nu+J}}\,.
\label{eq:intboundbulkbound}
\end{align}
This integrand can be written as the limit    of two bulk-to-boundary propagators
\be
 \int dP_2\, \frac{
\Pi_{h+i\nu,J}(X_2,P_2;W_2,D_{Z_2})
}{J! (h-1)_J }
\frac{1
}{ \mathcal{C}_{h-i\nu,J}}
\lim_{X_1\rightarrow P_1 \atop W_1 \rightarrow Z_1}
\Pi_{h-i\nu,J}(X_1,P_2;W_1,Z_2)
\,.
\ee
If one naively takes the limit outside,  
the   integral   is proportional to the harmonic function $\Omega_{\nu,J}(X_1,X_2;W_1,W_2)$, which itself can be written as a sum of two bulk-to-bulk propagators as in (\ref{integrationboundary}).
Then the limit $X_1 \to P_1$ of the bulk-to-bulk propagators just gives a sum of two bulk-to-boundary propagators from $X_2$ to $P_1$ with dimension $h+i\nu$ and $h-i\nu$. 
However, this cannot be correct because the original integral (\ref{eq:intboundbulkbound}) had dimension $h-i\nu$ at point $P_1$.
Dropping the term with wrong dimension, 
one obtains the   result
\footnote{The extra term was generated when we naively interchanged the limit with the integration symbol.}
\begin{align}
&\int dP_2\, \frac{\Pi_{h+i\nu,J}(X_2,P_2;W_2,D_{Z_2})(H_{12})^J }{J! (h-1)_J(P_{12})^{h-i\nu+J}}=-\frac{i}{2\nu\mathcal{C}_{h-i\nu,J}}\,
\Pi_{h-i\nu,J}(X_2,P_1;W_2,Z_1)\,.
\label{eq:IntegralBoundaryBoundaryWithBoundaryBulk} 
\end{align}

  The contribution from the term   proportional to $\d(\nu+\bar{\nu})$ in (\ref{eq:CompleteBulkIntegralSpinJ})  can be easily fixed  using a simple symmetry argument, 
  since the original integral (\ref{eq:StartingPoint})  is an even function of $\bar{\nu}$.
    Thus, we conclude that (\ref{eq:StartingPoint}) is given by  
\begin{align}
\frac{\big[\,\d\!\left(\nu+\bar{\nu}\right)+\d\!\left(\nu-\bar{\nu}\right)\big]\nu^2  }{2\pi J! (h-1)_J}
&\int dP_1 \, \Pi_{h+i\nu,J}(X_1,P_1;W_1,D_{Z_1})\,\Pi_{h-i\nu,J}(X_2,P_1;W_2,Z_1)\,, 
\end{align}
which shows (\ref{eq:ConstraintFinal}).


\subsection*{Completeness}
The goal of this section is to determine the coefficients  $c_{J,l}(\nu)$ that appear in  the completeness relation (\ref{eq:CompletenessNice}),  
\be
\sum_{l=0}^J\int d\nu \,c_{J,l}(\nu)\big( (W_1\cdot\nabla_1)(W_2\cdot\nabla_2)\big)^{l}\Omega_{\nu,J-l}(X_1,X_2;W_1,W_2) = 
\delta(X_1,X_2) (W_{12})^J\,.
\ee
Our  strategy will be to  find a recursion relation that fixes all coefficients $c_{J,l}(\nu)$ starting from the initial condition $c_{J,0}(\nu)=1$ derived in the main text.
With this in mind, we
take the divergence 
of the  equation  above at the point $X_1$.
On the right hand side we use 
\footnote{
To derive this formula one needs to use $W_1\cdot X_2 \,\delta(X_1,X_2)=0$ and similar identities.
}
\be
\nabla_1 \cdot K_1 
\left[\delta(X_1,X_2)(W_{12})^J\right]=
-\frac{J(2h+2J-3)}{2} \,W_2\cdot \nabla_2
\left[
\delta(X_1,X_2)(W_{12})^{J-1}
\right]\,, 
\ee
and on the left hand side we use the commutation relations
\begin{align}
\big[\nabla \cdot K,W\cdot\nabla \big]&=\bigg(\frac{2h-1}{2}+\mathcal{D}_{W}\bigg)\nabla^2 -\bigg( (\mathcal{D}_{W})^2+\frac{3(2h-1)}{2}\,\mathcal{D}_W +\frac{(2h-1)^2}{2}\bigg)\mathcal{D}_W\,,
\nonumber\\
\big[\nabla^2,W\cdot\nabla \big]&=-2 \big(h-1+\mathcal{D}_{W}\big)\,W\cdot \nabla \,,
\label{eq:BasicCommutators}
\end{align}
where $\mathcal{D}_W = W\cdot \partial_W$.
Using these basic commutators, one can show that
\be
\left[\nabla^2,(W\cdot\nabla_{X})^n\right]=-n (2h-1+2\mathcal{D}_{W}-n)(W\cdot \nabla_{X})^n\,.
\ee
Similarly, with a bit more effort, one finds
\begin{align}
\left[\nabla \cdot K,(W\cdot\nabla_{X})^l\right]=\ &
\frac{1}{2}(W\cdot\nabla_{X})^{l-1} 
 l ( 2 h + l + 2 \mathcal{D}_W-2 )\,\times \\
&\times 
\big[
1 - l - ( l + \mathcal{D}_W-1 ) ( 2 h + l + \mathcal{D}_W-2 ) +   \nabla^2
\big] \,.\nonumber
\end{align}
Using these commutators, the divergence
of equation (\ref{eq:CompletenessNice}) can be written as
\begin{align}
&
-W_2\cdot\nabla_2 
\sum_{l=1}^J\int d\nu \,c_{J,l}(\nu)\,
\frac{(l+1) (3-l-2 h-2 J ) \big[(h+J-1)^2+\nu^2\big]}{2}\, \times
\\
&
\big( (W_1\cdot\nabla_1)(W_2\cdot\nabla_2)\big)^{l-1}\Omega_{\nu,J-l}(X_1,X_2;W_1,W_2) 
=
\frac{J(2h+2J-3)}{2}\, W_2\cdot \nabla_2
\big[
\delta(X_1,X_2)W_{12}^{J-1}
\big]\,.
\nonumber 
\end{align}
Shifting the summation variable $l\to l+1$, we can identify this equation as $W_2\cdot \nabla_2$ acting on (\ref{eq:CompletenessNice}) with $J$ replaced by $J-1$.
This gives the recursion relation
\begin{align}
c_{J,l+1}(\nu)=
\frac{J (2 h+2 J-3)}{(l+1) (2 h+2
   J-l-3) } \frac{c_{J-1,l}(\nu)}{
   (h+J-1)^2+\nu
   ^2}\,,
\end{align}
which supplemented  with the initial condition $c_{J,0}(\nu)=1$ determines all $c_{J,l}(\nu)$.
In fact, one can write the general solution in closed form,
\be
c_{J,l}(\nu)=
\frac{2^l (J-l+1)_l
   \left(h+J-l-\frac{1}{2}\right)_l}
   {l! (2 h+2 J-2 l-1)_l (h+J-l-i
   \nu )_l (h+J-l+i \nu )_l}\,.
\ee

\section{Computation of partial amplitude}
\label{Ap:SplitRepCPW}
The goal of this appendix is to derive the  expression for the function $\alpha_l(\nu)$ appearing in the partial amplitude
(\ref{eq:SplitPartialWaveRelation}). The tensor operations present in (\ref{eq:Split4pt}) are of the form
\begin{align}
&\left(W_1\cdot\nabla_1\right)^{J-l}\Pi_{\Delta,l}=\mathcal{C}_{\Delta,l}\left(W_1\cdot\nabla_1\right)^{J-l}
\frac{\big( 2(W_1\cdot Z)(P_5\cdot X_1) -2(W_1\cdot P_5)(Z\cdot X_1)\big)^{l}}{\left(-2 P_5\cdot X_1\right)^{\Delta+l}}\nonumber\\
&=\mathcal{C}_{\Delta,l}\,\frac{\left(2P_5\cdot W_1\right)^{J-l}\big(2 (W_1\cdot Z)(P_5\cdot X_1)-(W_1\cdot P_5)(Z\cdot X_1)\big)^{l}\left(\Delta+l\right)_{J-l}}{\left(-2P_5\cdot X_1\right)^{\Delta+J}}\,,\label{eq:FirstTensorOperatorRel}\\
&\left(W_1\cdot\nabla_1\right)^J\Pi_{\Delta_2}=\frac{\mathcal{C}_{\Delta_2}\left(W_1\cdot\nabla_1\right)^J}{\left(-2P_2\cdot X_1\right)^{\Delta_2}}=\mathcal{C}_{\Delta_2}\left(\Delta_2\right)_{J}\frac{\left(2W_{1}\cdot P_2\right)^J}{\left(-2P_2\cdot X_1\right)^{\Delta_2+J}}\,,
\end{align}
and
\begin{align}
&(\star)\equiv \frac{\left(K\cdot \nabla\right)^J\Pi_{\Delta_2}\left(W\cdot\nabla\right)^{J-l}\Pi_{\Delta,l}}{\mathcal{C}_{\Delta_2}\mathcal{C}_{\Delta,l}J!\left(h-\frac{1}{2}\right)_{\!J}}
=
\label{eq:Kaction2}\\
&
=\left(\Delta_2\right)_{J}\left(\Delta+l\right)_{J-l}2^{J}\!\left(J-l\right)! \,\Gamma\!\left(3/2-h-J\right)
\frac{\big(2(P_2\cdot P_5)( X_1\cdot Z)-2(P_2\cdot Z)( P_5\cdot X_1)\big)^{l}}{\left(-2P_2\cdot X_1\right)^{\Delta_2+J}\left(-2P_5\cdot X_1\right)^{\Delta+J}}
\nonumber\\
&\ \ \ \times \sum_{m=0}^{J-l}\frac{\Gamma(1-h-J+m)\big(2(P_2\cdot X_1)( P_5\cdot X_1)\big)^{m}(P_{25})^{J-l-m}}
{\Gamma\!\left(1-h-J+\frac{m}{2}\right)\Gamma\!\left(\frac{3-2h-2J+m}{2}\right)m! \left(J-l-m\right)! }\,.
\nonumber
\end{align}
The first two operations follow almost immediately  from the definition, just notice that in (\ref{eq:FirstTensorOperatorRel}) the differential operator can act only in the denominator as it gives zero once it acts on the numerator. To obtain (\ref{eq:Kaction2}) we use (\ref{eq:ImportantTensorManipulation}).


The function $F_{\nu,J}$ is defined as the integral over the boundary of three-point functions. So, to derive $F_{\nu,J}$ from (\ref{eq:Split4pt}) we need to integrate over one of the bulk points, say $X_1$, generating a structure that has the form of a three-point function at points $P_1$, $P_2$ and $P_5$. Joining all the pieces that connect to the bulk point $X_1$, we have 
\begin{align}
&\int dX_1 \,\Pi_{\Delta_1} (\star)\,\mathcal{C}_{\Delta_2}  \, \mathcal{C}_{\Delta,l}=\frac{\mathcal{B}_{\Delta_1,\Delta_2,\Delta,l,J}\big((Z\cdot P_2)( P_1\cdot P_5)-(Z\cdot P_1)( P_2\cdot P_5)\big)^l}{P_{12}^{\frac{\Delta_2+l+\Delta_1-\Delta}{2}}P_{15}^{\frac{\Delta_1+\Delta+l-\Delta_2}{2}}P_{25}^{\frac{\Delta+\Delta_2+l-\Delta_1}{2}}}\,,
\end{align}
where the function $\mathcal{B}_{\Delta_1,\Delta_2,\Delta,l,J}$ is given by 
\begin{align}
\mathcal{B}_{\Delta_1,\Delta_2,\Delta,l,J}=\ &
\sum_{m=0}^{J-l}\frac{\mathcal{C}_{\Delta_2} \,\mathcal{C}_{\Delta,l}\left(\Delta_2\right)_J\left(\Delta+l\right)_{J-l}\left(-1\right)^{J+m}\left(J-l\right)!2^{1-2h-J}}{\sqrt{\pi}\left(J-l-m\right)!\,m!\,\mathcal{C}_{\Delta+J-m-l}}
\nonumber\\
&
\times
\frac{\Gamma\!\left(\frac{3-2h-2J}{2}\right)(m-h-J)!\left(\frac{\Delta+\Delta_1-\Delta_2-l}{2}\right)_{l} \,b_{\Delta_1,\Delta_2+J-m,\Delta+J-m-l,0}}
{\G(m+2-2h-2J)\left(\Delta+J-m-l\right)_l\,\mathcal{C}_{\Delta_2+J-m}}\,.
\end{align}
Notice that to perform the integration over $X_1$ we used an identity similar to (\ref{eq:DerivativeTrick3}). The function $b_{\Delta_1,\Delta_2,\Delta,J}$ is the same as in (\ref{eq:bfunction}). 
The integration over $X_2$ produces a similar term and so the next step  to read off the relation is to integrate the product of two three-point  functions over the boundary point. 

At this point we just need to bring the expression close to equation (3.16)  of \cite{DoMellin} that is related to the function $F_{\nu,J}$. So we just need to evaluate 
\begin{align}
&\int \!\!dP_5\, \frac{\big(P_{25}(D_Z\cdot P_1 )-P_{15}(D_Z\cdot P_2 )\big)^J}
{P_{15}^{\frac{\Delta_1+\Delta-\Delta_2+J}{2}}P_{25}^{\frac{\Delta_2+\Delta-\Delta_1+J}{2}}P_{12}^{\frac{\Delta_1+\Delta_2-\Delta+J}{2}}}
\frac{\big(P_{45}(Z\cdot P_3 )-P_{35}(Z\cdot P_4)\big)^J}
{P_{35}^{\frac{\Delta_3+d-\Delta-\Delta_4+J}{2}}P_{45}^{\frac{\Delta_4+d-\Delta-\Delta_3+J}{2}}P_{34}^{\frac{\Delta_3+\Delta_4-d+\Delta+J}{2}}}\,,
\end{align}
where the operator $D_Z$ is the projection operator (\ref{TodorovOper}), as defined in  \cite{SpinningCC,Costa:2011dw}. Its action produces a Gegenbauer polynomial. The integral over $P_5$, involving these polynomials, is precisely (3.16) of \cite{DoMellin}, where it was shown to be equal to a linear combination of conformal blocks. Taking into account the definition of the function $F_{\nu,J}$ in terms of conformal blocks\footnote{Notice that the conformal blocks of \cite{DoMellin} and \cite{Costa:2012cb} have different normalization.}, it is possible to extract the coefficient $\beta$,
\begin{align}
\beta_{\nu,\Delta_i,J}=\ &\frac{2^{3-2J}\pi^{1+h} \,\Gamma(i\nu)\,\Gamma(-i\nu)\,(h-i\nu-1)_J\,(h+i\nu-1)_J}
{\Gamma\!\left(\frac{\Delta_1+\Delta_2-h-i\nu+J}{2}\right)\Gamma\!\left(\frac{h+J+\Delta_1-\Delta_2+i\nu}{2}\right)\Gamma\!\left(\frac{h+J-\Delta_1+\Delta_2+i\nu}{2}\right)\Gamma\!\left(\frac{\Delta_1+\Delta_2+J+i\nu-h}{2}\right)}\nonumber\\
&\times \frac{1}{\Gamma\!\left(\frac{h+J+\Delta_3-\Delta_4-i\nu}{2}\right)\Gamma\!\left(\frac{h+J+\Delta_4-\Delta_3-i\nu}{2}\right)\Gamma\!\left(\frac{\Delta_3+\Delta_4+J-h-i\nu}{2}\right)\Gamma\!\left(\frac{\Delta_3+\Delta_4+i\nu-h+J}{2}\right)}\,.
\label{Beta}
\end{align}
The coefficient $\alpha_l(\nu)$ can be read after gathering all the components together, 
\begin{align}
\alpha_l=\mathcal{B}_{\Delta_1,\Delta_2,h+i\nu,l,J}\,\mathcal{B}_{\Delta_3,\Delta_4,h-i\nu,l,J}\,\frac{\nu^2 \beta_{\nu,\Delta_i,l}}{\pi J!(h-1)_J }\,,
\label{eq:ExplicitEquationAlpha}
\end{align}
or explicitly,
\be
\a_l(\nu)=\frac{((J-l)!)^2  \G^2 \!\left(\frac{3-2h-2J}{2}\right) R_{J,l}(\nu,\D_1,\D_2)\,R_{J,l}(-\nu,\D_3,\D_4)}{\pi ^{3 h+1}2^{4 h+4 J+2 l+3}  \G (\Delta _1+1-h)\G (\Delta _2+1-h) \G (\Delta _3+1-h) \G (\Delta_4+1-h) }\,,
\ee
where we defined
\begin{align}
R_{J,l}(\nu,\D_1,\D_2) =
\sum_{p=0}^{J-l} \ &\frac{(-1)^p\G (p+1-h-J) (h+J-p+i \nu )_p \left(J-p+\Delta _2\right)_p }{(J-l-p)!p!\G (p+2-2h-2J)}
\nonumber\\
& \times\left(\frac{\Delta _1+\Delta _2-h+l+i \nu}{2}\right)_{J-l-p}\left(\frac{h+l+i \nu -\Delta _{12}}{2}\right)_{J-l-p}\,.
\end{align}

\bibliographystyle{./utphys}
\bibliography{./mybib}

\end{document}